\documentclass[useAMS,usenatbib]{mn2e}
\usepackage{times}
\usepackage{graphicx}

\usepackage{journals}

\title[Transient dynamics in Keplerian discs]
{A study of transient dynamics of perturbations in Keplerian discs using a variational approach}
\author[V. V. Zhuravlev and D. N. Razdoburdin]{V. V. Zhuravlev$^{1}$\thanks{E-mail:
zhuravlev@sai.msu.ru} and D. N. Razdoburdin$^{1,2}$\\
$^{1}$Sternberg Astronomical Institute, Moscow M.V. Lomonosov State University, Universitetskij pr., 13, Moscow 119992, Russia\\
$^{2}$Department of Physics, Moscow M.V. Lomonosov State University, Moscow, 119992, Russia}
\begin{document}

\date{Accepted
. Received ; in original form }

\pagerange{\pageref{firstpage}--\pageref{lastpage}} \pubyear{2013}

\maketitle

\label{firstpage}

\begin{abstract}

We study linear transient dynamics in a thin Keplerian disc employing a method based on 
variational formulation of optimisation problem.
It is shown that in a shearing sheet approximation due to a prominent excitation of density waves by vortices
the most rapidly growing shearing harmonic has azimuthal wavelength, $\lambda_y$, of order of the disc thickness, $H$, 
and its initial shape is always nearly identical to a vortex having the same potential vorticity.
Also, in the limit $\lambda_y\gg H$ the optimal growth $G\propto (\Omega/\kappa)^4$, 
where $\Omega$ and $\kappa$ stand for local rotational and epicyclic frequencies, respectively, 
what suggests that transient growth of large scale vortices can be much stronger in areas with non-Keplerian rotation, 
e.g. in the inner parts of relativistic discs around the black holes. 
We estimate that if disc is already in a turbulent state with effective viscosity given by 
the Shakura parameter $\alpha<1$, the considered large scale vortices with wavelengths $H/\alpha>\lambda_y>H$ 
have the most favourable conditions to be transiently amplified before they are damped. 
At the same time, turbulence is a natural source of the potential vorticity for this transient activity. 
We extend our study to a global spatial scale showing that global perturbations with azimuthal wavelengths 
more than an order of magnitude greater than the disc thickness still are able to attain the growth of dozens 
of times in a few Keplerian periods at the inner boundary of disc.

\end{abstract}


\section{Introduction}

A conventional way to study non-stationary phenomena in accretion discs is spectral (i.e. modal) analysis
when disc eigen-frequencies are determined by looking for the set of solutions
that vary exponentially with time, see e.g. \citet{K2001} for basic review. The next problem is whether the modes
are really excited due to spectral instabilities, turbulent motions or external forcing.
However, as we show in this work, a prominent transient growth of amplitudes of global vortices with azimuthal 
wavelengths larger than the disc thickness can be obtained employing the non-modal approach to the dynamics
of small perturbations in geometrically thin discs.

Contrary to the modal framework the non-modal approach sorts out
perturbations according to the amount of energy they gain from the background during a specified time interval,
see  \citet{S2007} and \citet{SH2001}.
Mathematically, the set of eigen-vectors of underlying dynamical operator
is replaced by the set of its singular vectors. The key point is that
the latter is time-dependent and strongly differs from the former in case of shear, i.e. non-normal, flow.
Projection of an arbitrary initial perturbation onto eigen-vectors gives information
about its long-term behaviour, while the projection onto singular vectors additionally represents
its potential to transient dynamics, i.e. disc response shortly after some abrupt event triggered perturbation.
Clearly, this response can be quite different from which one expects from spectral analysis.
The component corresponding to the highest singular value is usually called an optimal perturbation.
In previous works, the study of transient dynamics was concentrated mainly
on local analysis made in the shearing sheet approximation, see e.g. \citet{Ch2003}.
A few investigations have been devoted to global transient dynamics including
a real accretion disc geometry in order to reveal an additional mechanism for enhanced angular momentum transfer
\citep{Um2006,R2009, S2010}. The global perturbations have been also examined for the transient dynamics by \citet{IK2001},
which remains a unique study of the global optimal growth in Keplerian discs made so far. However, \citet{IK2001}
considered the problem of angular momentum transport restricting their analysis to incompressible perturbations.
This is quite a strong restriction if one addresses the issue of the non-stationary appearance of geometrically thin disc due
to the transient effects. Thus, here we would like to tackle the optimal configurations of
compressible perturbations and corresponding optimal growth factors. Along with this task we
would like to illustrate a potential of a relatively new technique of the optimisation which
is based on the variational framework and has been successfully applied
to a number of complex hydrodynamical flows \citep{LB1998,CB2001,G2006}.
The advantage of the variational approach is that in contrast to the usual optimisation method
it does not rely on the representation in the basis of modes or on any other discretisation procedure.
This implies that it can be easily generalised to non-stationary background flows and
even to non-linear problem.

This paper is organised as follows. First we describe a general formalism of the optimisation method. It shows 
that optimal perturbation can be determined by using an iterative loop in which the set of basic dynamical equations
is integrated forward in time and the set of corresponding adjoint equations is integrated backward in time. 
It is discussed that this method is applicable to perturbations in stationary as well as non-stationary flows. 
Next we consider the dynamics of the perturbed rotating flow specifying the sets of equations for linear perturbations 
on a global spatial scale as well as in the shearing sheet approximation.
We also discuss a choice of norm to measure the transient growth in case of compressible fluid.
It is suggested to use the norm which equals to the canonical energy of perturbations in axisymmetric case.  
For all options we derive the adjoint equations. Then, we present the optimal growth calculations in 
geometrically thin Keplerian disc comparing the results in different approaches. 
Additionally, in the Appendix A a problem of the non-modal growth of optimal 
axisymmetric perturbations measured by their acoustic energy is studied analytically in detail. 
At last, in the Appendix B we supply the description of the optimisation procedure in application to a model 
case of incompressible perturbations in global and local context.

\section{Variational formulation of optimisation problem}

\subsection{Method of Lagrange multipliers}

In order to solve an optimisation problem one has to find initial conditions that excite the most powerful 
perturbation at a given time interval, $\tau$. While dealing with a linear problem for stationary background flow 
a common strategy is to consider a linear subspace of solutions of dynamical equations 
which is represented by linear span of a finite number of modes. 
By modes we mean spectral solutions with exponential dependence on time $\propto e^{{\rm i}\sigma t}$.
Then, it is necessary to determine the coefficients in the combination of modes that maximise 
the growth of perturbation. This can be done, for example, by means of the singular value decomposition of the relevant 
matrix exponential projected onto the orthonormal basis, look \citet{SH2001}.
\citet{BF1992} use this method to study the optimal transient growth in classical Couette, Poiseuille and 
Blasius boundary layer flows, \citet{Nar2005} use it to investigate Keplerian flow in local approximation and 
\citet{ZhSh2009} and, subsequently, \citet{RZh2012} apply it to study the global optimal growth in 
quasi-Keplerian torus with free boundaries. 
However, this strategy entails all technical difficulties related to calculation of modes and eigen-frequencies. 
Among them, complications that one encounters solving the boundary problem in the vicinity of 
singular points such as corotational or Lindblad resonances. Further, since the modes are non-orthogonal 
a high dimension of their linear span may be required to obtain the reliable results. 
Also, the contribution of the continuous spectrum to transient dynamics remains unclear. The latter is particularly
important when the unbounded flows are considered. At last, this method can not be generalised in order to
investigate the behaviour of perturbations in non-stationary flows as well as 
the behaviour of perturbations with finite amplitude. 

Instead, the optimisation problem may be alternatively formulated in terms of a variational principle.
For a brief introduction to this subject see
\citet{S2007}, a more detailed exposition can be found in book by \citet{Gunz2003}.
Below we would like to give a generic view on the variational framework which will be applied to concrete 
astrophysical flow later on.

So, all what we need to do is to maximise the functional

\begin{equation}
\label{cost}
\mathcal{G}(\tau) = \frac{||{\bf q}(\tau)||^2}{||{\bf q}(0)||^2},
\end{equation}
which is defined in the space of perturbation state vectors, ${\bf q}$. Each element ${\bf q} = \{q^i(t)\}$
consists of the set of perturbation quantities evolving with time\footnote{Say, perturbations of pressure
and velocity components as functions of time and spatial coordinates.}. $\mathcal{G}$ is usually 
called an objective, or cost, functional of the problem.
By eq. (\ref{cost}) we imply that the inner product is defined in this functional space,

\begin{equation}
\label{inner}
({\bf q_1},{\bf q_2}) = \int M_{ij} {\rm Re} [\,q_1^i \bar q_2^j\,]\, {\rm d}V,
\end{equation}
which is real for an arbitrary pair of vectors, overbar indicates complex conjugation
and $M_{ij}$ is a certain real, symmetric and positive definite matrix,
so that the norm $||{\bf q}||=\sqrt{({\bf q},{\bf q})}$ characterises amplitude of perturbations.
This can be either the kinetic energy in case of incompressible dynamics, or the acoustic energy
if one includes a finite sound speed, or it can be any other positive definite physically motivated quantity.

The key ingredient of the method is that the maximum of $\mathcal{G}$ we are looking for is
the conditional one, since ${\bf q}$ is constrained by the requirement that it obeys the basic dynamical equations,
\begin{equation}
\label{sys}
\frac{\partial {\bf q}}{\partial t} = {\bf A}\, {\bf q}
\end{equation}
The system (\ref{sys}) contains a dynamical (differential) operator ${\bf A}$ that controls the evolution of perturbation
quantities.
This suggests that actually we have to implement a constrained optimisation, what can be done by
the Lagrangian multipliers method. The latter is the generalised version of finding the conditional extrema
of functions when the Lagrangian multiplier emerges as the proportionality factor between the gradients of 
an objective function and constraint function. In the calculus of variations functionals replace functions
whereas the Lagrangian multipliers become functions themselves. 
In other words, we formally change to the extended space of
vectors without the restriction (\ref{sys}) and introduce an additional
so called adjoint vectors, $\tilde {\bf q}$, which will serve as the Lagrangian multipliers.
After that, one can define what is usually called an augmented Lagrangian involving both ${\bf q}$ and $\tilde{\bf q}$
in the following way
\begin{equation}
\label{augmented}
\mathcal{L} ({\bf q},\tilde{\bf q}) = \mathcal{G}({\bf q}) -
\int \limits_0^\tau ( \, \tilde {\bf q}\, ,  \dot {\bf q} - {\bf A} {\bf q} \,)\, {\rm d}t,
\end{equation}
where the partial time derivative is denoted by dot. The second term in eq. (\ref{augmented}) is called penalty
term, i.e. it penalises the objective Lagrangian, $\mathcal{G}$, each time when ${\bf q}(t)$ does not conform 
eq. (\ref{sys}).

All what remains is to find an unconditional extremum of $\mathcal{L}$ which is given
by the zero variations of $\mathcal{L}$ with respect to an arbitrary variations of both ${\bf q}$ and $\tilde{\bf q}$,
e.g. in the case of the adjoint vector one has to require the vanish of
\begin{equation}
\delta \mathcal{L} = \lim\limits_{\epsilon\to 0} 
\frac{\mathcal{L}({\bf q},\tilde {\bf q}+\epsilon \delta \tilde {\bf q}) - L({\bf q},\tilde {\bf q})}{\epsilon} = 0
\end{equation}
for an arbitrary function $\delta \tilde {\bf q}$.

Obviously, the zero variation of $\mathcal{L}$ with respect to $\tilde{\bf q}$ recovers the system (\ref{sys}),
whereas to vary $\mathcal{L}$ over ${\bf q}$ one has to integrate the second term in eq. (\ref{augmented})
by parts. This reads
\begin{equation}
\label{by_parts}
\int \limits_0^\tau ( \, \tilde {\bf q}\, ,  \dot {\bf q} - {\bf A} {\bf q} \,)\, {\rm d}t =
(\tilde{\bf q},{\bf q})\bigr|^\tau_0 -
\int \limits_0^\tau ( \,\dot {\tilde {\bf q}} + {\bf A}^\dag \tilde {\bf q}, \, {\bf q} \,)\, {\rm d}t,
\end{equation}
where we use the ordinary definition of the adjoint operator, 
$(\tilde {\bf q},{\bf A}{\bf q}) = ({\bf A}^\dag \tilde {\bf q}, {\bf q})$, through the inner
product (\ref{inner}). It is now straightforward to see, that the variation of $\mathcal{L}$ with respect to an arbitrary
deviation $\delta {\bf q}$ gives
the set of so called adjoint equations,
\begin{equation}
\label{adj_sys}
\frac{\partial \tilde {\bf q}}{\partial t} = - {\bf A}^\dag \, \tilde {\bf q},
\end{equation}
and the following additional relations

\begin{equation}
\label{edge_terms1}
\tilde {\bf q}(\tau) = \frac{2}{||{\bf q}(0)||^2} \, {\bf q}(\tau),
\end{equation}
\begin{equation}
\label{edge_terms2}
{\bf q}(0) = \frac{||{\bf q(0)}||^4}{2\,||{\bf q(\tau)}||^2} \, \tilde {\bf q}(0).
\end{equation}
Indeed, at first take arbitrary $\delta {\bf q}$ that vanish in the neighbourhood of $t=0$ and $t=\tau$. Then, 
only the second (`volume') term in eq.~(\ref{by_parts}) contributes to $\delta \mathcal{L}$ what implies 
eq. (\ref{adj_sys}). Once we have that eq.~(\ref{adj_sys}) holds inside the interval $(0,\tau)$ we see 
that the remaining `edge' terms in $\delta \mathcal{L}=0$ imply that
\begin{equation}
2{\bf q}(\tau)\,\frac{1}{||{\bf} q(0)||^2} - 2{\bf q}(0)\,\frac{||{\bf q}(\tau)||^2}{||{\bf q}(0)||^4} - 
\tilde {\bf q} \, \Bigr|^\tau_0 = 0,
\end{equation}
what gives eq. (\ref{edge_terms1}) and eq. (\ref{edge_terms2}) since $\delta {\bf q}$ can vanish 
at $t=0$ and at $t=\tau$ independently.

Additionally, we assume here that both $\tilde {\bf q}$ and ${\bf q}$ satisfy appropriate boundary
conditions. Note that for physical problem considered below, where we derive an
explicit form of ${\bf A}^\dag$, the boundary conditions for the adjoint variables
are obtained using together the zero variation of $\mathcal{L}$ and the boundary conditions
for the state variables. This is shown using the integration
by parts in the spatial domain, i.e. similarly to what was done above with a time dependence.

Thus, eqs. (\ref{sys},\ref{adj_sys}) are coupled through the conditions (\ref{edge_terms1},\ref{edge_terms2})
and must be solved together. The unique solution gives
both the state and the adjoint vectors that correspond to a maximum of
the cost functional (\ref{cost}), $G\equiv \max\{\mathcal{G}\}$, provided that dynamical equations (\ref{sys}) are
satisfied. Physically, this means that we find an optimal initial perturbation attaining the highest
possible energy growth at a given time interval, $\tau$. $G(\tau)$ itself is usually called the optimal growth. 
Besides, while considering below the evolution of the particular perturbation, no matter optimised or not, we characterise 
it by growth factor, $g(t) \equiv ||{\bf q}(t)||^2 / ||{\bf q}(0)||^2$.

\subsection{Operator solutions and iterative scheme of optimisation}

At least in the linear case, natural method of solution of these coupled sets of equations 
stems from the fact that the optimal state vector is
the first singular vector of propagator, ${\bf U}$, that advances perturbations up to $t=\tau$,
namely, ${\bf q}(\tau) = {\bf U}{\bf q}(0)$; for a short but clear account look also \citet{L2000}.

An explicit form of ${\bf U}$ can be obtained solving eq. (\ref{sys}).
At first, let us suppose that ${\bf A}$ is independent of time (an autonomous operator, see \citet{FI1996a} for reference).
Then ${\bf U} = e^{{\bf A}t}$ what can be seen from eq. (\ref{sys}). 
From the operator theory it is known that the first singular value of any operator is the square root of the largest
eigenvalue of a positive definite composite operator which is the original times its adjoint.
Thus, in order to solve an optimisation problem one has to determine the largest eigenvalue of
${\bf U}\cdot{\bf U}^\dag = e^{{\bf A}t} \cdot e^{{\bf A}^\dag t}$ which is equivalent
to the advance of perturbation first forward in time using eq. (\ref{sys}) and
then {\it backward} in time using  eq. (\ref{adj_sys}), because of the minus appearing in eq. (\ref{adj_sys}).
The direct way to converge to the largest eigenvalue of ${\bf U}\cdot{\bf U}^\dag$
is to iterate an arbitrary initial perturbation advancing it
recurrently by the operator itself,
$(\, {\bf U}\cdot{\bf U}^\dag \,)^{p\to\infty}\, {\bf q}(0)$, where $p$ is a natural number.
This procedure is usually called the power iteration, see e.g. \citet{Golub1996} for details. 
It can be shown that the power iteration is equivalent to a steepest descent algorithm for finding an extremum
of $\mathcal{L}$, look e.g. \citet{Gunz1994}.

In a more general case when ${\bf A}$ depends on time (non-autonomous operator, see \citet{FI1996b} for reference)
${\bf U}$ may be represented as an ordered product of infinitesimal propagators 
\begin{equation}
\label{U_t}
{\bf U}(t) = \lim\limits_{n\to \infty} \prod \limits_{j=1}^n e^{{\bf A}(t_j)\delta t}, 
\end{equation}
where $\delta t=t/n$ and $(j-1)\delta t < t_j < j\delta t$.
Then, according to the rule of taking the adjoint of composite operator, the adjoint of propagator, 
${\bf U}^\dag$, is given by
\begin{equation}
\label{adj_U_t}
{\bf U}^\dag (t) = \lim\limits_{n\to \infty} \prod \limits_{j=n}^1 e^{{\bf A}^\dag(t_j)\delta t}, 
\end{equation}
where the adjoint infinitesimal propagators stand in the reverse order, i.e. to advance 
some initial vector one takes ${\bf A}^\dag(t_j)$ starting from the final point of time interval and moving back 
to $0$. Again, the action of $e^{{\bf A}^\dag(t_j)\delta t}$ is equivalent to the integration of the system (\ref{adj_sys})
backward in time from $j\delta t$ to $(j-1)\delta t$. Thus, according to eq.~(\ref{adj_U_t}) the action of ${\bf U}^\dag$
is identical to the integration of the system~(\ref{adj_sys}) {\it backward} in time. 
We see that the action of ${\bf U}\cdot{\bf U}^\dag$ is equivalent to forward and backward advance of perturbation 
solving eq.~(\ref{sys}) and eq.~(\ref{adj_sys}), respectively, just like in the case of ${\bf A}$ independent of time.

It should be noted that the existence of the largest eigenvalue of ${\bf U}\cdot{\bf U}^\dag$ is 
guaranteed by the Krein-Rutman theorem of functional analysis, see \citet{KR1950}. For the details on the
related subject of compact operators the reader is referred to \citet{Kolm1961}.

As a result, no matter whether the system (\ref{sys}) contains coefficients dependent on time or not, 
the underlying optimisation problem is naturally solved 
integrating the basic and the adjoint dynamical equations,
(\ref{sys}) and (\ref{adj_sys}), forward and backward in time, respectively,
with the conditions (\ref{edge_terms1},\ref{edge_terms2})
linking the state and the adjoint vectors at the turning points of the loop.
Note that we have not done any assumptions about the background flow. So, the
iteration scheme described above can be employed in a wide class of complex flows
when the solving of the spectral problem commonly used to evaluate
the transient growth can be a quite involved task.
Moreover, as it is shown above, there is no stationarity restriction of the background. Hence, there is
no technical obstacles to investigate the transient dynamics that may be triggered
in the non-stationary accretion discs or other types of astrophysical shearing flows, say, jets and winds.
The situation is more complicated if one tries to apply an iterative loop to a non-linear problem, but
nevertheless a number of technical improvements have been devised for this case, see sect. 6 of the review by Schmid (2007)
and references therein.

\section{Perturbations in rotating shear flow}

To apply the general formalism described in the previous section in astrophysical context, we would like to
consider small perturbations in a disc, i.e. in axisymmetric rotating flow.
If one neglects the effects of viscosity and consider only the model case of baratropic
equation of state, then the dynamics of small perturbations is described by the set of linear equations

\begin{equation}
\label{orig_sys1}
\frac{\partial \delta {\bf v}}{\partial t} + ({\bf v}\cdot\nabla)\delta {\bf v} + (\delta {\bf v}\cdot\nabla) {\bf v} = 
-\nabla \delta h,
\end{equation}

\begin{equation}
\label{orig_sys2}
\frac{\partial \delta\rho}{\partial t} + \nabla \cdot(\rho \delta{\bf v}) + \nabla \cdot (\delta \rho {\bf v}) = 0,
\end{equation}
where $\delta {\bf v}$ and
$\delta h$ are the Eulerian perturbations of velocity and enthalpy, $\delta \rho$ is the Eulerian perturbation
of density.
In our case $\delta h=\delta p/\rho$, where $\delta p$ is the Eulerian perturbation of pressure and $\rho(r,z)$ 
is the background density.
We will use the cylindrical coordinates $(r,\varphi,z)$ in which the background flow is described 
by azimuthal motion ${\bf v} = (0,v_\varphi,0)$ with angular velocity $\Omega = v_\varphi / r$
that depends on the radial coordinate only.

In this study we also assume that perturbations preserve vertical hydrostatic equilibrium. 
In case of baratropic flow the vertical hydrostatic equilibrium results
in perturbations with no dependence on $z$, what makes possible to integrate the dynamical equations along the vertical
direction, see e.g. \citet{Gold1986}. Let us note that in general the assumption
of absence of vertical motions in the perturbed flow can be strictly justified only if $t_{pert}\gg \Omega^{-1}$ and
$\lambda_{pert}\gg H$, with $t_{pert}$ and $\lambda_{pert}$ being the characteristic time and length of perturbations and 
$H$ being the thickness of disc. However, even without this restriction 
still there are particular cases when perturbations with no node in vertical 
direction can exist in the flow, see for example \citet{Okaz87} who 
showed that vertical and planar perturbed motions can be separated from each other
in thin disc with isothermal vertical structure. So in this study we would like not to restrict 
ourselves with the above rigorous assumption about $t_{pert}$ and $\lambda_{pert}$.
An additional argument in favour of the model case of vertically independent perturbations in the context of 
non-modal analysis comes from 
the study by \citet{Yecko04} who investigated the transient growth of 3D local incompressible perturbations in 
viscous Keplerian shear. The largest growth factors were found for perturbations uniform along the z axis. 
Thus, in what follows we consider the planar perturbed velocity field, $\delta {\bf v} = (\delta v_r,\delta v_\varphi,0)$,
and work with the set of equations (\ref{orig_sys1}, \ref{orig_sys2}) integrated over $z$. 
Since the background flow is rotationally symmetrical we are dealing with the azimuthal Fourier harmonic of
perturbations $\propto e^{{\rm i} m\varphi}$ hereafter\footnote{In section 3.2 devoted to 
local dynamics we start considering perturbations with general dependence on $\varphi$.} implying that perturbation quantities are functions of 
time and radial coordinate.

Below in this section we discuss the specific equations that have to be solved in order to determine 
the optimised perturbations. We measure perturbations using two different norms and derive 
the specific adjoint equations for both of them.  Along with the full optimisation problem 
which is formulated to investigate global perturbations and 
accurately accounts for the cylindrical geometry as well as for the shear rate distribution across the flow, 
we study its local spatial limit using the well-known shearing sheet approximation.    

Although our objective is to consider transient dynamics in hypersonic flow, a complementary description
of the optimisation problem in model case of incompressible fluid can be found in the Appendix B. 
There we give necessary equations constructed for perturbations of vorticity and stream function.
Numerical tests carried out using these equations allowed us to make an additional check of our 
primary numerical scheme for compressible dynamics. 
Apart, in the shearing sheet limit the variational procedure for the divergence-free velocity perturbations becomes
especially simple and can be performed fully analytically. Hence, we obtain an exact analytical expression
for optimal growth as function of time in this case.

\subsection{Optimisation on a global spatial scale}

Let us assume that the state vector ${\bf q}$ is constructed from
the Eulerian perturbations of the velocity components and the enthalpy,
${\bf q} = \{ \delta v_r,\,\delta v_\varphi,\,\delta h\}$. 
The explicit form of $\bf A$ in eq. (\ref{sys}) is
\begin{equation}
\label{A_expl}
\left ( \begin{array}{ccc}
-im\Omega & 2\Omega & - \partial_r \\
-\kappa^2/(2\Omega) &  -im\Omega & -im/r \\
-a_*^2 \left (\, (r\Sigma)^{-1} \partial_r(r\Sigma) + \partial_r \right ) & -a_*^2\, im/r  & -im\Omega
\end{array} \right )
\end{equation}
where $\Sigma=\int \rho dz$ is the surface density and $a_*^2=na_{eq}^2/(n+1/2)$ with $a_{eq}$ being the sound speed
in the equatorial plane of the flow, $n$ is the polytropic index. $\Sigma$ and $a_{eq}$ both have a
specified dependence on $r$ and
$\kappa^2 = (2\Omega/r) d/dr (\Omega r^2)$ is the epicyclic frequency squared. 

\subsubsection{Choice of norm and adjoint equations}

It is not a matter of course, what norm of compressible perturbations to choose in order to measure their growth 
appropriately. 
The very first idea that comes to mind is to choose $M_{ij}$ in eq. (\ref{inner}) in such a way that 
the norm of each state vector equals to
the total acoustic energy of perturbation which
reads

\begin{equation}
\label{ac_en}
{||{\bf q}||^2}_1 = \pi \int \Sigma \left ( |\delta v_r|^2 + |\delta v_\varphi|^2  + \frac{|\delta h|^2}{a_*^2}  \right ) r\, dr
\end{equation}
In eq. (\ref{ac_en}) it is implied that integration over the azimuthal and vertical coordinates is done.
The variant to measure perturbations by their acoustic energy seems to be somewhat natural since this
quantity is physically meaningful and it is conserved in absence of shear.

However, it turns out that the norm (\ref{ac_en}) leads to indication of non-modal growth of axisymmetric 
compressible perturbations. The latter has an oscillatory rather than the transient behaviour. 
This is not difficult to show in the shearing sheet approximation, what is done in present work and 
is relegated to the Appendix A. 

Strictly speaking, the general oscillatory solution can also be referred to as a non-modal growth, see e.g. 
section 4.2 in paper by \citet{Afsh2005} who considered an axisymmetric perturbations in incompressible limit
but including the vertical motions. In spite that, one would like to exclude this particular case 
reserving the pure transient dynamics. Fortunately, this is possible to do since the basic equations (\ref{sys}) 
with $\bf{A}$ given by eq. (\ref{A_expl}) allow for an energy-like integral in case $m=0$. 


Indeed, eq. (\ref{A_expl}) taken with $m=0$ yields 
$$
\delta v_r \frac{\partial (\delta v_r)}{\partial t} + 
\frac{4\Omega^2}{\kappa^2} \delta v_\varphi \frac{(\partial \delta v_\varphi)}{\partial t} + 
\frac{\delta h}{ a_*^2} \frac{\partial (\delta h)}{\partial t} = 
\frac{1}{r\Sigma} \frac{\partial (r\Sigma\delta v_r \delta h)}{\partial r},
$$
what leads to conservation of the following quantity
\begin{equation}
\label{E_c}
E_c = \pi \int \Sigma \left ( |\delta v_r| ^2 + \frac{4\Omega^2}{\kappa^2}|\delta v_\varphi|^2  + \frac{|\delta h|^2}{a_*^2}  \right ) r\, dr,
\end{equation}
provided that $\Sigma$ vanishes at the boundaries of the flow.

A further inspection reveals that eq. (\ref{E_c}) is nothing but the canonical energy in the particular case of 
axisymmetric perturbations in $(r\varphi)$ domain. 
It is not difficult to verify this fact looking at the general expression for $E_c$ 
derived by \citet{FS1978} (FS hereafter) in the Lagrangian framework concerning linear perturbations
settled in the rotating axially symmetric flow (see their eq. (45)~).

Let us first note that the 6th term in square brackets in eq. (45) by FS yields
\begin{eqnarray}
\label{can_expr1}
|\xi_r|^2 (\partial_{rr}p + \rho\partial_{rr}\Phi)+ 
\frac{|\xi_\varphi|^2}{r} (\partial_r p + \rho \partial_r\Phi) = \nonumber \\ 
\rho\left [ \Omega^2 (|\xi_r|^2+|\xi_\varphi|^2) + 2\Omega r\partial_r\Omega |\xi_r|^2 + 
\frac{1}{a^2\rho^2}(\partial_r p)^2 |\xi_r|^2 \right ],
\end{eqnarray}
where we keep the notations of FS.
The first term in square brackets in eq. (\ref{can_expr1}) cancels the geometric terms coming from 
$-\rho |v\cdot \nabla\xi|^2$ which enters eq. (45) of FS.
At the same time, the last term therein along with the rest of thermal terms entering  eq.(45) of FS 
gives the thermal contribution to $E_c$, $\Sigma|\delta h|^2/ a_*^2$, including the usual change to 2D polytropic index, $n \to n+1/2$,
after the integration over the disk thickness.
At last, with the help of kinematic relation (11) by FS and explicit form of $\xi_\varphi$
derived by \citet{FS1978_2} directly below their eq. (21) we find that
$$
|\xi_r|^2 + |\xi_\varphi|^2 + 2\Omega r\partial_r\Omega |\xi_r|^2 = 
|\delta v_r|^2 + \frac{4\Omega^2}{\kappa^2} |\delta v_\varphi|^2,
$$ 
what confirms that eq. (\ref{E_c}) is the canonical energy of axisymmetric compressible (baratropic) perturbations
expressed in the Eulerian variables. 
Note that $E_c$ is positive definite what allows us to use it as a norm for our optimisation
method. 

Thus, the definition 
\begin{equation}
\label{ac_en2}
{||{\bf q}||^2}_2 = E_c
\end{equation}
for norm of an arbitrary perturbation (including those with non-zero $m$) 
itself excludes the non-modal growth of axisymmetric perturbations. 

In this work it is instructive to use both norms, eq. (\ref{ac_en}) and eq. (\ref{ac_en2}), in equal rights 
in order to obtain a broader picture of transient dynamics. 
Also, the variant (\ref{ac_en2}) is used for the first time in the context of non-modal approach in discs 
in contrast to more familiar eq. (\ref{ac_en}), so one would like to 
trace the difference in results for optimal growth as well as for shapes of optimal perturbations.

Once the inner product is defined, we can derive the explicit form of ${\bf A}^\dag$. This is done
by writing the penalty term in eq. (\ref{augmented}) with help of explicit form of eq. (\ref{inner}) and
${\bf A}$ given by eq. (\ref{A_expl}). Using the equalities
$
\delta \tilde h \nabla\cdot (\Sigma\delta {\bf v}) = \nabla\cdot (\delta \tilde p \delta {\bf v} ) -
\Sigma \delta {\bf v}\cdot \nabla \delta \tilde h
$ and
$
\Sigma \delta \tilde {\bf v}\cdot \nabla \delta  h =
\nabla\cdot ( \delta p\, \delta\tilde {\bf v} ) - \delta h \nabla\cdot ( \Sigma \delta\tilde {\bf v} )
$ in the $r$-domain,
implementing the variation of the final expression over arbitrary deviations of
$\delta {\bf v}$ and $\delta h$ and setting the result to zero we obtain from the edge term
that $\delta \tilde p=0$ at the boundary provided that $\delta p=0$ ibidem. The volume integral gives the adjoint
set of equations so that ${\bf A}^\dag$ takes the form

\begin{equation}
\label{A_adj_expl}
\left ( \begin{array}{ccc}
im\Omega & -\kappa^2/(2\Omega) & \partial_r \\
2\Omega &  im\Omega & im/r \\
a_*^2 \left (\, (r\Sigma)^{-1} \partial_r(r\Sigma) + \partial_r \right ) & a_*^2\, im/r  & im\Omega
\end{array} \right )
\end{equation}
in case of norm given by eq.(\ref{ac_en}) and

\begin{equation}
\label{A_adj_expl2}
\left ( \begin{array}{ccc}
im\Omega & -2\Omega & \partial_r \\
\kappa^2/(2\Omega) &  im\Omega  &  \frac{im}{r} \,\frac{\kappa^2}{4\Omega^2} \\
a_*^2 \left (\, (r\Sigma)^{-1} \partial_r(r\Sigma) + \partial_r \right ) & a_*^2 \frac{im}{r}\, \frac{4\Omega^2}{\kappa^2} & im\Omega
\end{array} \right )
\end{equation}
in case of norm given by eq. (\ref{ac_en2}).

Looking at eq. (\ref{A_adj_expl}) we see that it slightly differs from eq. (\ref{A_expl}). Apart the opposite
sign which actually annihilates with minus in eq. (\ref{adj_sys}) when one writes the set of equations,
it has two off-diagonal elements flipped over. Namely, $\kappa^2/(2\Omega)$ in the second raw
takes place of $2\Omega$ in the first raw and vice verso. This stands for the fact that
for a rigidly rotating flow we must have ${\bf A} = -{\bf A}^\dag$,
consequently ${\bf U} \cdot {\bf U}^{\dag} = {\bf I}$ (since for rigid rotation
$\kappa = 2\Omega$), and no transient growth effects because there is no shear in that case.
A different thing happens while changing to the inner product according to eq. (\ref{ac_en2}): 
in eq. (\ref{A_adj_expl2}) we find another two off-diagonal terms multiplied (divided) by the factor 
$4\Omega^2/\kappa^2$ which differs from unit in the presence of shear.

\subsubsection{Implementation of iterative loop}

To test the variational technique and the underlying iterative procedure for determination of global optimal
perturbations we implement a numerical integrator of the Cauchy problem for the basic and the
adjoint equations. Since ${\bf A}$ and ${\bf A}^\dag$ are given by 
eqs. (\ref{A_expl}) and (\ref{A_adj_expl}) (or eq. (\ref{A_adj_expl2}) if changing to different norm), respectively, we follow 
\citet{FR1988} and choose a leap-frog scheme since this is a simple explicit second-order method which is 
stable for wave-like dynamics. Four different meshes with constant
coordinate and time steps, $\Delta r$ and $\Delta t$, are introduced on the $(r,t)$-plane. 
The second and the third meshes are shifted for $\Delta t/2$ along the time axis and for $\Delta r/2$ along 
coordinate axis relatively to the first one. The fourth mesh is shifted both in time and space for 
$\Delta t/2$ and $\Delta r/2$. Then, we split the basic and the adjoint sets of equations into real and 
imaginary parts and assign real and imaginary parts of $\delta v_r,\, \delta v_\varphi,\, \delta h$ to corresponding
meshes in such a way that for a particular equation approximations of time and spatial derivatives 
are centred at the same nodes. 
In order to advance the large scale perturbations with $\lambda_{pert}$ comparable to the radial scale of
variations of background quantities we have to impose the boundary conditions. 
We require that the Lagrangian perturbation of enthalpy vanishes at the inner boundary of the flow. 
Note that if $\Sigma \to 0$ at the boundary, it is sufficient to impose the regularity condition
on the perturbed quantities therein. 
At the same time, the outer boundary condition is of no concern, since the outer boundary is assumed to 
be located far beyond the radial domain occupied by perturbations evolved until $t=\tau$. 

\begin{figure}
\begin{center}
\vspace{1cm}
\includegraphics[width=8.5cm,angle=0]{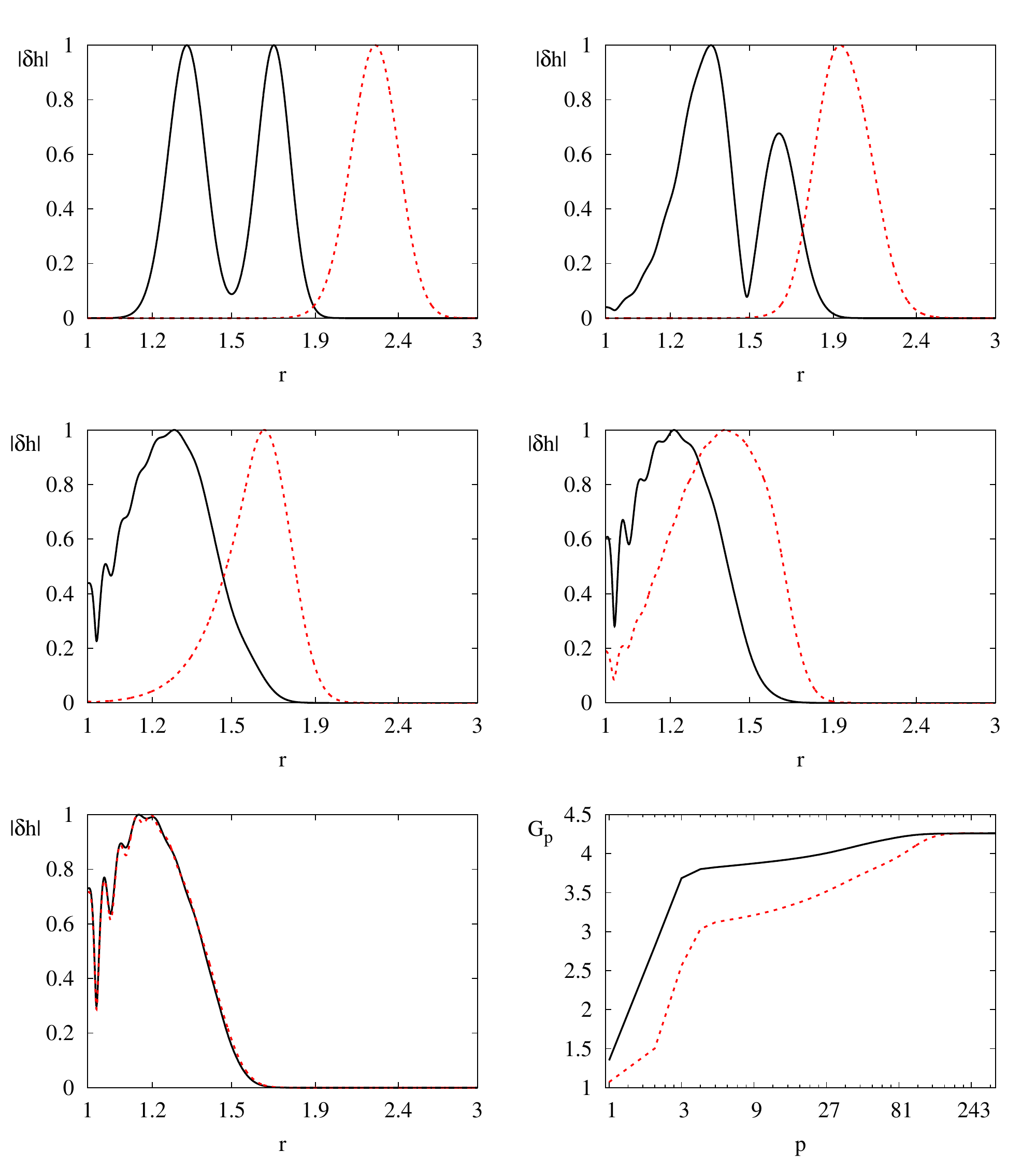}
\end{center}
\caption{Convergence of an arbitrary initial profiles of $|\delta h|$ to an optimal initial profile in Keplerian
disc with structure specified by 
eqs. (\ref{sig}, \ref{a_eq}). 
It is implied that perturbations are measured according to eq. (\ref{ac_en}). 
In the left- and right-hand top panels, left- and right-hand middle panels and left-hand bottom panel 
the initial profiles of $|\delta h|$ are presented corresponding to 
$p=0,25,80,120$ and $350$ iterations, respectively.
In the right-hand bottom panel there is an iterational value of optimal growth, $G_p(\tau)$, vs. the number of iteration. 
The dashed and solid styles for curves in panels are used for two different iterative loops that 
launch with 'single' and 'double' Gaussian functions
given by eqs. (\ref{gauss1}) and (\ref{gauss2}).
$\tau=3, m=5, n=1.5, \delta=0.05$. } \label{fig1}
\end{figure}

Regarding the background flow, we use mainly a Keplerian disc with uniform distribution of $\Sigma$ and 
$a_{eq}$ what should be suitable to make comparisons with local dynamics in order to see how 
the distribution of the shear rate and the cylindrical geometry affects transient growth. 
Thus, we set
\begin{equation}
\label{homogen_disc}
\Sigma=const, \quad a_{eq} = \delta / \sqrt{2n} \quad \mbox{and} \quad \Omega = r^{-3/2},
\end{equation}
which we will refer to as 'homogeneous' disc model hereafter. In the global approach 
it is assumed that $r$ is given in units of inner radius of disc and all time intervals 
are measured in units of inverse Keplerian frequency at $r=1$. The constant $\delta \sim H/r \ll 1$
specifies an aspect ratio of geometrically thin disc.
Also, we would like to check another variant of the background flow
which is specified by the following profiles
\begin{equation}
\label{sig}
\Sigma \propto  r^{-3/5} (1-r^{-1/2})^{3/5},
\end{equation}
\begin{equation}
\label{a_eq}
a_{eq} = (\delta/\sqrt{2n})\,\Omega\,  r^{21/20} (1-r^{-1/2})^{1/5},
\end{equation}
$\Omega$ takes its Keplerian value as well. 
Eqs. (\ref{sig}) and (\ref{a_eq}) are adopted from \citet{SS73} 
as a representative of the thin accretion disc model. 

Now, in order to find an optimal perturbation we specify an arbitrary initial state vector, i.e. we take 
an arbitrary initial condition, ${\bf q}_{0}(t=0)$, 
and integrate the basic equations forward in time up
to some $t=\tau$. Then we substitute the result, ${\bf q}_0(t=\tau)$, as the initial condition to 
the adjoint equations (depending on norm chosen for perturbations) 
and integrate them backward in time up to 
$t=0$ getting the next variant of the initial state vector, ${\bf q}_1(t=0)$. 
At last, ${\bf q}_1(t=0)$ must be renormalised, i.e. divided by its own norm calculated according 
to eq. (\ref{ac_en}) (alternatively, to eq. (\ref{ac_en2}) ). The first iteration is accomplished now. 
The iterative loop consists of a number of such iterations necessary to achieve the desirable accuracy 
of determination of the optimal vector ${\bf q}_{opt}(t)=\lim\limits_{p\to\infty} {\bf q}_{p}(t)$ and 
corresponding optimal growth 
\begin{equation}
\label{G_gl}
G(\tau) \equiv ||{\bf q}_{opt}(\tau)||^2 / ||{\bf q}_{opt}(0)||^2 = 
\lim\limits_{p\to\infty} G_p(\tau),
\end{equation}
where $G_p(\tau)\equiv||{\bf q}_{p}(\tau)||^2 / ||{\bf q}_{p}(0)||^2$.

Foremost, let us check a convergence of this iterative loop. 
In the case of norm given by eq. (\ref{ac_en}) and background profiles given by 
eqs.(\ref{sig}), (\ref{a_eq})  it is illustrated in fig. (\ref{fig1}).
To launch iterations we take two distinct starting shapes of ${\rm Re}[\delta h]$ as the initial condition for the Cauchy problem:
the 'single' and the 'double' Gaussian functions, which have arbitrary positions and radial dispersions, 
\begin{equation}
\label{gauss1}
{\rm Re}[\delta h] \bigr |_{t=0} \propto \exp( \,-(\sqrt{r}-\mu_1)^2/(2\sigma_1^2)\, ),
\end{equation}
\begin{eqnarray}
\label{gauss2}
{\rm Re}[\delta h]\bigr |_{t=0} \propto \exp(\,-(\sqrt{r}-\mu_2)^2/(2\sigma_2^2)\,) + \nonumber \\ 
  \exp(\,-(\sqrt{r}-\mu_3)^2/(2\sigma_3^2)\,),
\end{eqnarray}
where $\mu_{1,2,3}$ and $\sigma_{1,2,3}$ are arbitrary numbers. The rest of the quantities constructing 
the state vector are set to zero at $t=0$. 
Using our numerical scheme after
$\sim 10^2$ iterations we get a unique optimal initial shape of perturbation with the optimisation timespan,
$\tau=3$. At the same time, the iterational value of optimal growth, $G_p(\tau)$,
converges to the highest possible value, $G(\tau)$, as seen in the bottom right-hand panel in fig. (\ref{fig1}).
Let us stress that we obtain the same optimal initial shape of $|\delta h|$ independently on the particular
values of $\mu_{1,2,3}$ and $\sigma_{1,2,3}$. 
A quite similar situation takes place in case of norm given by eq. (\ref{ac_en2}).

\subsection{Optimisation in a shearing sheet model}

In the local spatial limit the evolution of compressible perturbations can be considered in the
shearing sheet approximation. The corresponding equations have been derived by \citet{GL1965}, see 
also the paper by \citet{UR2004} for a detailed account of this derivation. The case of the linear 
perturbations in the context of non-modal approach was studied by \citet{Bodo05}, B05 hereafter. 
Equations that we need can be quoted from B05. 
Those are the following
\begin{equation}
\label{sonic_sys1}
\left ( \frac{\partial}{\partial t} - q\Omega_0 x\frac{\partial}{\partial y} \right ) u_x - 2\Omega_0 u_y =
-\frac{\partial W}{\partial x},
\end{equation}
\begin{equation}
\label{sonic_sys2}
\left ( \frac{\partial}{\partial t} - q\Omega_0 x\frac{\partial}{\partial y} \right ) u_y + 
(2 - q)\Omega_0 u_x =
-\frac{\partial W}{\partial y},
\end{equation}
\begin{equation}
\label{sonic_sys3}
\left ( \frac{\partial}{\partial t} - q\Omega_0 x\frac{\partial}{\partial y} \right ) W + 
a_*^2 \left ( \frac{\partial u_x}{\partial x} + \frac{\partial u_y}{\partial y} \right ) = 0,
\end{equation}
where $u_x,u_y$ are the Eulerian perturbations of the velocity components and 
$W$ is the Eulerian perturbation of enthalpy excited in a small patch of the disc, 
$x \equiv r-r_0, y \equiv r_0(\varphi-\Omega_0 t)$ are the local Cartesian coordinates, $x,y\ll r_0$, that 
correspond to reference frame rotating with the angular velocity $\Omega_0 = \Omega(r_0)$.  
$q \equiv - (r/\Omega) (d \Omega/dr )|_{r=r_0}$ is
the constant shear rate that defines the background velocity as $v^{loc}_y=-q\Omega_0 x$ and
all other background quantities such as $\Sigma$ and $a_*$ are assumed to be constant.

Let us employ an iterative procedure elucidated in the previous section.
In order to do this we have to find equations that are adjoint to the set
(\ref{sonic_sys1}-\ref{sonic_sys3}). Evidently, the shearing sheet limit of eq. (\ref{A_adj_expl}) 
is reproduced by the following set of equations

\begin{equation}
\label{adj_sonic_sys1}
\left ( \frac{\partial}{\partial t} - q\Omega_0 x\frac{\partial}{\partial y} \right ) \tilde u_x - (2 - q)\Omega_0\tilde u_y =
-\frac{\partial\tilde W}{\partial x},
\end{equation}
\begin{equation}
\label{adj_sonic_sys2}
\left ( \frac{\partial}{\partial t} - q\Omega_0 x\frac{\partial}{\partial y} \right ) \tilde u_y + 
2\Omega_0 \tilde u_x =
-\frac{\partial \tilde W}{\partial y},
\end{equation}
\begin{equation}
\label{adj_sonic_sys3}
\left ( \frac{\partial}{\partial t} - q\Omega_0 x\frac{\partial}{\partial y} \right ) \tilde W + 
a_*^2 \left ( \frac{\partial \tilde u_x}{\partial x} + \frac{\partial \tilde u_y}{\partial y} \right ) = 0
\end{equation}
or, alternatively, the shearing sheet limit of eq. (\ref{A_adj_expl2}) is 
reproduced by the another set of equations,
\begin{equation}
\label{adj_sonic_sys1_2}
\left ( \frac{\partial}{\partial t} - q\Omega_0 x\frac{\partial}{\partial y} \right ) \tilde u_x - 2\Omega_0\tilde u_y =
-\frac{\partial\tilde W}{\partial x},
\end{equation}
\begin{equation}
\label{adj_sonic_sys2_2}
\left ( \frac{\partial}{\partial t} - q\Omega_0 x\frac{\partial}{\partial y} \right ) \tilde u_y + 
(2 - q)\Omega_0 \tilde u_x =
-\frac{2 - q}{2}\,\frac{\partial \tilde W}{\partial y},
\end{equation}
\begin{equation}
\label{adj_sonic_sys3_2}
\left ( \frac{\partial}{\partial t} - q\Omega_0 x\frac{\partial}{\partial y} \right ) \tilde W + 
a_*^2 \left ( \frac{\partial \tilde u_x}{\partial x} + \frac{2}{2-q} 
\frac{\partial \tilde u_y}{\partial y} \right ) = 0.
\end{equation}

In eqs. (\ref{adj_sonic_sys1}-\ref{adj_sonic_sys3_2})  
it is assumed that $\tilde u_x,\tilde u_y$ and $\tilde W$ stand for the adjoint velocity and enthalpy perturbations.

Finally, we would like to change to the dimensionless comoving Cartesian coordinates,
$x^\prime = \Omega_0 x/a_*,\, y^\prime = \Omega_0 (y+q\Omega_0 xt)/a_*,\, t^\prime=\Omega_0 t$
\footnote{Due to vertical hydrostatic equilibrium in disc this implies that we express lengths in terms of 
disc thickness, $H = a_*/\Omega_0$}, 
what leads us to spatially homogeneous set of equations
that is, however, inhomogeneous in time. Nevertheless, considering partial solutions in the form
of shearing harmonic (or, more accurately, spatial Fourier harmonic, abbreviated SFH)
$
f = \hat f (k_x,k_y,t^\prime) \exp ({\rm i} k_x x^\prime + {\rm i} k_y y^\prime),
$
where $f$ is any of the unknown quantities, $\hat f$ is its Fourier amplitude and 
$(k_x,k_y)$ are the dimensionless wavenumbers along $x$ and $y$ axes expressed in units of $\Omega_0/a_*$,
we obtain the corresponding set of ODEs

\begin{equation}
\label{sonic_sys1_sh}
\frac{d \hat u_x}{d t^\prime} = 2\hat u_y - {\rm i}\,(k_x+k_y q t^\prime) \hat W,
\end{equation}

\begin{equation}
\label{sonic_sys2_sh}
\frac{d \hat u_y}{d t^\prime} = -(2 - q) \hat u_x - {\rm i}\, k_y \hat W,
\end{equation}

\begin{equation}
\label{sonic_sys3_sh}
\frac{d \hat W}{d t^\prime} = - {\rm i}\, ( \, (k_x+k_y q t^\prime) \hat u_x + k_y \hat u_y \,),
\end{equation}

\begin{equation}
\label{adj_sonic_sys1_sh}
\frac{d \hat{\tilde u}_x}{d t^\prime} = 
(2 - q) \hat{\tilde u}_y - {\rm i}\, (k_x+k_y q t^\prime) \hat {\tilde W},
\end{equation}

\begin{equation}
\label{adj_sonic_sys2_sh}
\frac{d \hat {\tilde u}_y}{d t^\prime} = 
- 2 \hat {\tilde u}_x - {\rm i}\, k_y \hat {\tilde W},
\end{equation}

\begin{equation}
\label{adj_sonic_sys3_sh}
\frac{d \hat {\tilde W}}{d t^\prime} = 
- {\rm i}\, ( \, (k_x+k_y q t^\prime ) \hat {\tilde u}_x + k_y \hat {\tilde u}_y \,),
\end{equation}
where 
eqs. (\ref{sonic_sys1_sh}-\ref{sonic_sys3_sh}) have to be solved in order to determine the state vector,\\
${\bf q}(t^\prime)=\{\hat u_x(t^\prime), \hat u_y(t^\prime), \hat W(t^\prime)\}$, and 
eqs. (\ref{adj_sonic_sys1_sh}-\ref{adj_sonic_sys3_sh}) have to be solved in order to determine the adjoint vector, 
${\bf \tilde q}(t^\prime)=\{\hat{\tilde u}_x(t^\prime), \hat {\tilde u}_y(t^\prime), \hat{\tilde W}(t^\prime)\}$. 
Eqs. (\ref{sonic_sys1_sh}-\ref{adj_sonic_sys3_sh}) contain the adjoint part which results from eqs. 
(\ref{adj_sonic_sys1}-\ref{adj_sonic_sys3}), whereas in case of eqs. (\ref{adj_sonic_sys1_2}-\ref{adj_sonic_sys3_2}) 
it has to be replaced by the following set of equations

\begin{equation}
\label{adj_sonic_sys1_sh_2}
\frac{d \hat{\tilde u}_x}{d t^\prime} = 
2 \hat{\tilde u}_y - {\rm i}\, (k_x+k_y q t^\prime) \hat {\tilde W},
\end{equation}
\begin{equation}
\label{adj_sonic_sys2_sh_2}
\frac{d \hat {\tilde u}_y}{d t^\prime} = 
- (2 - q) \hat {\tilde u}_x - \frac{2 - q}{2} {\rm i}\, k_y \hat {\tilde W},
\end{equation}

\begin{equation}
\label{adj_sonic_sys3_sh_2}
\frac{d \hat {\tilde W}}{d t^\prime} = 
- {\rm i}\, ( \, (k_x+k_y q t^\prime) \hat {\tilde u}_x + 
\frac{2}{2 - q} k_y \hat {\tilde u}_y \,).
\end{equation}

Throughout eqs. (\ref{sonic_sys1_sh}-\ref{adj_sonic_sys3_sh_2}) it is implied that SFH of velocity perturbations and 
SFH of enthalpy perturbations are expressed in units of $a_*$ and $a_*^2$, respectively.
We omit the prime after $t$ hereafter.
Seeking for a solution to eqs. (\ref{sonic_sys1_sh}-\ref{adj_sonic_sys3_sh}) we 
use the surface density of the acoustic energy of a single
SFH,
\begin{equation}
\label{loc_ac_en}
E = \frac{1}{2S}\int\limits_S \left ( ({\rm Re} [u_x])^2 + ({\rm Re} [u_y])^2 + \frac{({\rm Re} [W])^2}{a_*^2} \right ) \,  {\rm d} x {\rm d}y,
\end{equation}
as a norm for local perturbations. 
Eq. (\ref{loc_ac_en}) leads to the following expression for SFH
\begin{equation}
\label{loc_norm}
{||{\bf q}||^2}_1 = \frac{1}{2} \left ( |\hat u_x|^2 + |\hat u_y|^2 + |\hat W|^2  \right ), 
\end{equation}
what is the local analogue of eq. (\ref{ac_en}).

On the other hand, in order to find a local counterpart of optimal perturbations measured by eq. (\ref{ac_en2}) we 
have to solve eqs. (\ref{sonic_sys1_sh}-\ref{sonic_sys3_sh}, \ref{adj_sonic_sys1_sh_2}-\ref{adj_sonic_sys3_sh_2})
employing the norm which reads
\begin{equation}
\label{loc_norm_2}
{||{\bf q}||^2}_2 = \frac{1}{2} \left ( |\hat u_x|^2 + \frac{2}{2 - q}\,|\hat u_y|^2 + |\hat W|^2  \right ).
\end{equation}

\subsubsection{On the relevant parametrisation of the problem}

Two types of shearing harmonics are allowed to exist in compressible medium. Those are vortices and 
density waves, look \citet{Ch1994}, \citet{Ch1997a} and B05.
Both are decoupled from each other in case when the perturbed motion is
subsonic, i.e. when the difference of shear velocities on the length-scale of the problem 
is less than sound speed.
The relevant length-scale in the shearing sheet is defined by the wavelength of SFH, 
$\lambda_{x} \sim H |k_x+k_y qt |^{-1}$, across the shear.
Thus, the condition that vortices and density waves live separately in the shearing box reads
\begin{equation}
\label{decoupled}
\lambda_x q\Omega_0 / a_* = \frac{q}{|k_y(k_x/k_y + qt )|} = \frac{R}{|\beta + qt |} < 1, 
\end{equation}
where we introduce the new parameters, $R$ and $\beta$, expressed through the usual wavenumbers as
$R=q/|k_y|$ and $\beta = k_x/k_y$. 
At least for Keplerian shear, $R$ is of order of $\lambda_y/H$ characterising azimuthal 
scale of SFH relative to the disk thickness. At the same time, for the shearing harmonic with $\beta<0$ 
the latter parameter defines 
the time of swing, $t _s = -\beta/q$, 
i.e. the instant when SFH changes its form from leading to trailing spiral. 
Previous studies have shown that during this event vortices stop gaining energy from the background 
and switch to decay phase, whereas density waves exhibit exactly the opposite behaviour.
The initially leading spirals are of particular interest in this study since their vortical configurations 
are subject to transient growth. 
 
Eq. (\ref{decoupled}) leads to apparent but important conclusion worth to be discussed here. 
We see that the leading spirals always pass a period when vortex motion is inseparable from acoustic motion.
This 'swing interval' is confined by the instants 
\begin{equation}
\label{coupled_interval}
t _{s_{1,2}} = -(\beta/q)\, (1\pm R/\beta), 
\end{equation}
with $t _{s_1}<t _s<t _{s_2}$. 
It is small compared to a characteristic evolution time of the leading spiral given by $t _s$ if 
\begin{equation}
\label{coupled_cond}
R\ll |\beta|/2,
\end{equation}
thus, not necessarily in the case of small azimuthal wavelengths, $R\ll 1$.
Generally, the sufficiently tightly wound (either leading or trailing) spirals with $\lambda_y\gg H$ ($R\gg 1$) can still
be a well-defined vortex or density wave. Though, it must be noted that the swing interval becomes short
with respect to the dynamical timescale, $(t _{s_2}-t _{s_1})\ll 1$, 
only if SFH is truly small scaled comparing to the disc thickness, i.e.
if $\lambda_y\ll H$ ($R\ll 1$).
Also note that eq. (\ref{decoupled}) is modified when $a_*$ becomes comparable or less than 
$\kappa (\lambda_x^2+\lambda_y^2)^{1/2}$ (where $\kappa^2 = 2(2-q)\Omega_0^2$ in the local case) 
since epicyclic oscillations become significant in the last case.
However, it can be checked that correction to estimate (\ref{coupled_interval}) is always of order of unity. 

Further, \citet{Ch1997b} and B05 described a phenomenon of generation of density waves by vortices 
as they swing from leading to trailing spirals. This process is asymmetric in a sense that vortices
are able to excite density waves but not vice verso. 
Later on, \citet{HP09_1} (HP hereafter) developed an analytical theory of density wave excitation within 
WKBJ framework. They obtained analytical expressions for amplitude and phase of density wave that emerges at 
swing time of vortex, $t _s$. The amplitude of density wave is proportional to $\epsilon^{-1/2}\exp(-4\pi/\epsilon)$ 
(see eq. (53) by HP), where $\epsilon$ is assumed to be a small parameter of the theory,  
\begin{equation}
\label{epsilon}
\epsilon = \frac{q k_y}{k_y^2 + \kappa^2/\Omega_0^2} = \frac{R}{1 + 2(2-q) R^2/q^2}.
\end{equation}
From eq. (\ref{epsilon}) we see that density wave excitation is suppressed in the limit of small azimuthal 
wavelengths, $R\ll 1$ (since $\epsilon\sim R$), as well as in the limit of large azimuthal wavelengths, 
$R\gg 1$ (since $\epsilon\sim R^{-1}$ if $q$ is not too close to $2$). 

Summarising this section, we expect that outside the swing interval defined by eq. (\ref{coupled_interval}) 
an arbitrary initial SFH can always be represented as combination of vortex and density wave and 
if the vortex constituting a part of SFH is a leading spiral it generates an additional density wave at the swing 
time, $t _s$. However, the latter event is substantial only if $R \sim 1$.  
Bearing in mind the general picture briefly exposed in this section we regard the parameters $\beta$ and $R$ 
as suitable to make a subsequent analysis of optimal SFH and we use them below to present our results.

\subsubsection{Transient growth of vortices in compressible medium}

As has been discussed by HP and by others, the vortical perturbations in compressible shear flow can be 
recognised as the slowly evolving solutions with non-zero potential vorticity. 
Indeed, eqs. (\ref{sonic_sys1_sh}-\ref{sonic_sys3_sh}) can be reformulated as the 2nd-order inhomogeneous 
equations for $\hat u_y$ (see eq. (32) by B05 or, alternatively, eq. (23) by HP), 
$\hat u_x$ and $\hat W$ (see eq. (22) by HP). The RHS of these equations are proportional to 
SFH of the potential vorticity being a time invariant. 
In our notations and in a form defined by B05 the latter quantity reads
\begin{equation}
\label{vorticity}
I = (k_x+k_y q t ) \hat u_y  - k_y \hat u_x + {\rm i} (2-q) \hat W
\end{equation}
(note that $I = -{\rm i}\hat\zeta$, where $\hat\zeta$ is the same quantity but as defined by HP).
Then, the vortical solutions are obtained if one discards the second time derivatives in equations.
In our notations these solutions are the following (see eq. (31) by HP) 
\begin{equation}
\label{vort_u_x}
\hat u^\prime_x = - \frac{K + 2q}{K^2 + 4 q^2 k_y^2} \, k_y I,
\end{equation}
\begin{equation}
\label{vort_u_y}
\hat u^\prime_y =  \frac{k_x + k_y q t }{K} \, I,
\end{equation}
\begin{equation}
\label{vort_W}
\hat W^\prime = 2{\rm i} \, \frac{q k_y^2 - K}{K^2 + 4 q^2 k_y^2} \, I,
\end{equation}
where
$K \equiv k_y^2+(k_x + k_y q t )^2 + \kappa^2/\Omega_0^2$. It is implied that eqs. (\ref{vorticity}-\ref{vort_W})
are given for dimensionless quantities.

As follows from the reasoning in section 3.2.1, eqs. (\ref{vort_u_x}-\ref{vort_W}) are 
a good approximation to an accurate solution of eqs. (\ref{sonic_sys1_sh}-\ref{sonic_sys3_sh}) obtained for 
some initial vortex perturbation if two conditions are satisfied: it is considered outside the swing interval
and its azimuthal wavelength significantly differs from the disc thickness (i.e. $\epsilon\ll 1$).
Despite of such strict limitations to solution (\ref{vort_u_x}-\ref{vort_W}) we would like to 
check to what estimations of growth factors it leads.
Using eq. (\ref{loc_norm}) we find for the acoustic energy of vortex that 
\begin{equation}
\label{SFH_E_compr}
\frac{2E}{a_*^2 I^2} = \frac{(k_x+k_y q t )^2}{K^2} + \frac{K_1}{K^2 + 4q^2 k_y^2}
\end{equation}
with $K_1 = 4 + k_y^2$.

Eq. (\ref{SFH_E_compr}) provides us with an expression for $g$ as function of $k_x, k_y$ and $t $.
In order to obtain the local analogue of $G$ introduced by eq. (\ref{G_gl}) we have to find 
maximum of $g(k_x)$ for fixed $k_y$ and $t $. To avoid straightforward but cumbersome
calculations redundant in our current estimations we assume that it is close 
to the growth factor of SFH swinging at $t $. Thus, we approximate the optimal growth 
by $G\approx g(k_x = - k_y q t )$ what yields

\begin{eqnarray}
\label{G_compr}
G \approx 
\frac{K_1 \left [K_0 + (k_y qt )^2 \right ]^2}{\left [K_1 + (k_y qt )^2 \right ]
\left [K_0 + (k_y qt )^2 \right ]^2 + 4 (k_y^2 q^2 t )^2} 
\nonumber \\
\frac{\left [K_0 + (k_y qt )^2 \right ]^2 + 4q^2 k_y^2}{K_0^2 + 4q^2 k_y^2}
\end{eqnarray}
with $K_0 = K(t  = -k_x/(qk_y))$.

Furthermore, eq. (\ref{G_compr}) should be considered in the limits of small ( $k_y \ll 1$ ) and
large ($k_y\gg 1$) azimuthal wavenumbers when excitation of density waves is suppressed and 
non-modal growth is represented solely by vortices.

Particularly, in the limit $k_y \ll 1$ ( strictly, as long as $R \gg q\Omega_0/\kappa$) we obtain that
\begin{equation}
\label{G_highR}
G \approx \frac{4\Omega_0^4}{\kappa^4} \frac{\left ( q^4 {t }^2/R^2 + \kappa^2/\Omega_0^2\right )^2}
{\left ( q^4 {t }^2/R^2 + 4 \right )},
\end{equation}
where $k_y$ is replaced by $R$.

Note that for a sufficiently long $t $ eq. (\ref{G_highR}) becomes especially simple. 
Strictly, in the case $t\gg 2R/q^2$
\begin{equation}
\label{G_highRt}
G \approx \frac{4\Omega_0^4}{\kappa^4} \frac{q^4 {t }^2}{R^2},
\end{equation}
what demonstrates that transient growth drops down inversely to the second power of $R$, whilst 
it remains constant for constant ratio $t /R$. Another important point is that 
$G$ rapidly increases as the background flow tends to constant angular momentum shear. 
Indeed, given that $\kappa^2 = 2 (2-q)\Omega_0^2 $ we see that $G\propto q^4 / (2-q)^2 \to \infty$ as $q\to 2$.
This result indicates that transient growth of large scale vortices may be of primary importance 
close of the last stable orbit in relativistic discs around the black holes. 
However, this is not the case in the opposite limit. 
For $k_y\gg 1$ ( strictly, as long as $R \ll q/2$) eq. (\ref{G_compr}) yields 
\begin{equation}
\label{G_lowR}
G \approx 1 + {(qt )}^2
\end{equation}
what is consistent with basic conclusions of \citet{Afsh2005} who also treated analytically the more realistic
cases adding viscosity and vertical component of small scale vortices. 
See also the results by \citet{Yecko04} in this context.
Note that this case is equivalent to $a_*\to\infty$, i.e. to the limit of incompressible dynamics when 
the velocity perturbation takes a divergence-free form. There exists a simple analytical solution
of the corresponding initial value problem and it becomes possible to obtain an exact 
analytical expression for $G(t )$, see eq. (\ref{SFH_G}) and the rest of the Appendix B for details.   
\\

Finally, let us assess the influence of non-zero viscosity which may effectively emerge 
through turbulent motions in disc. 
If it were not for unwinding due to the shear the initially tightly wound leading spiral would disperse on 
the timescale $\Delta t_\nu \sim \lambda_x^2 / \nu$, where $\nu$ is a kinematic viscosity coefficient. 
We employ its usual parametrisation through the Shakura $\alpha$-parameter, $\nu = \alpha a_* H$, 
getting that the dimensionless $\Delta t_\nu \sim (\alpha k_x^2)^{-1}$ rapidly decreases as $k_x$ becomes 
larger. At the same time, it takes a longer time for the transient growth to occur since 
$\Delta t_{tg} \sim |k_x/(q k_y)|$.
From the other hand, while the spiral unwinds its radial scale-length increases 
allowing the viscous dispersal to be suspended. 
Thus, the condition $\Delta t_{tg} = \Delta t_\nu$ puts a lower limit on the longest duration of non-modal growth
of vortex in viscous flow. 
Using the latter equality we obtain that 
\begin{equation}
\label{T_max}
\max (\Delta t_{tg}) \gtrsim \alpha^{-1/3} \left (\frac{R}{q^2} \right )^{2/3}
\end{equation}
It can be checked that eq. (\ref{T_max}) recovers an estimate given by \citet{Afsh2005}, see their eq. (81).
The upper bound on optimal growth corresponding to $\max(\Delta t_{tg})$
is given by its inviscid value, $G_{max}$, which is as follows
\begin{equation}
\label{G_max}
G_{max} \approx \frac{4\Omega_0^4}{\kappa^4} \left ( \frac{q^2}{\alpha R} \right )^{2/3}.
\end{equation}
Estimate (\ref{G_max}) is obtained for the large scale vortices ($R\gg 1$) 
by substituting eq. (\ref{T_max}) into eq. (\ref{G_highRt}).
We see that according to this approximate expression 
the transient growth ceases only for vortices with $H/\lambda_y \sim \alpha$ which, in turn, 
becomes a marginal condition for moderately viscous discs with $\delta \sim \alpha$.

\subsubsection{Implementation of iterative loop}

In the shearing sheet model we construct a local counterpart of the scheme described in section 3.1.2. 
This time, an arbitrary initial condition, ${\bf q}_{0}(t =0)$,
consists of a single SFH of velocity and enthalpy perturbations and is used to 
integrate eqs. (\ref{sonic_sys1_sh}-\ref{sonic_sys3_sh}) forward in time. 
Again, we use the result, ${\bf q}_0(t =\tau)$, as the initial condition to 
eqs. (\ref{adj_sonic_sys1_sh}-\ref{adj_sonic_sys3_sh}) 
(alternatively, to eqs. (\ref{adj_sonic_sys1_sh_2}-\ref{adj_sonic_sys3_sh_2}) ) and integrate them backward in time. 
After ${\bf q}_1(t =0)$ is renormalised, i.e. divided by its own norm given by 
eq. (\ref{loc_norm}) (alternatively, by eq. (\ref{loc_norm_2}) ), it is used 
in the next iteration. 

Note that in this way $G_p$ and $G$ are determined for particular values of $k_x$ and $k_y$, or, alternatively, 
for particular values of $\beta$ and $R$.
Thus, we imply hereafter that the local optimal growth is the quantity
obtained in iterative loop for single SFH. However, we denote it further explicitly as $G(\beta)$ since, by default, 
$G$ is defined as the optimal growth for the specified azimuthal wavenumber ($k_y$ in local or $m$ in global context).  

Besides, the numerical parametrical study of optimal growth carried out with the help of standard 
GNU Scientific Library routine for integration of sets of ODEs shows that there always exists 
some value of $\beta = \beta_{max}(p)$ where $G_p(\beta)$ 
attains an absolute maximum provided that the other parameters (including $R$) are fixed. 
Consequently, renormalising ${\bf q}_p(t , \, \beta)$ by the norm of ${\bf q}_p(t , \, \beta=\beta_{max}(p))$
for each iteration we eventually get the optimal state vector ${\bf q}_{opt}(t , \beta=\beta_{max})$
corresponding to a single SFH. This result is independent on the initial state vector, 
${\bf q}_{0}(t , \, \beta)|_{R=const}$,
which can be any packet of shearing harmonics. At the same time, the optimal growth corresponding to 
optimal vector obtained in this way, $G$, is equal to absolute maximum of $G(\beta)$ mentioned previously.

\section{Optimal solutions in a Keplerian disc}

\subsection{Inspection of optimal perturbations in shearing sheet model}

\begin{figure}
\begin{center}
\includegraphics[width=8.5cm,angle=0]{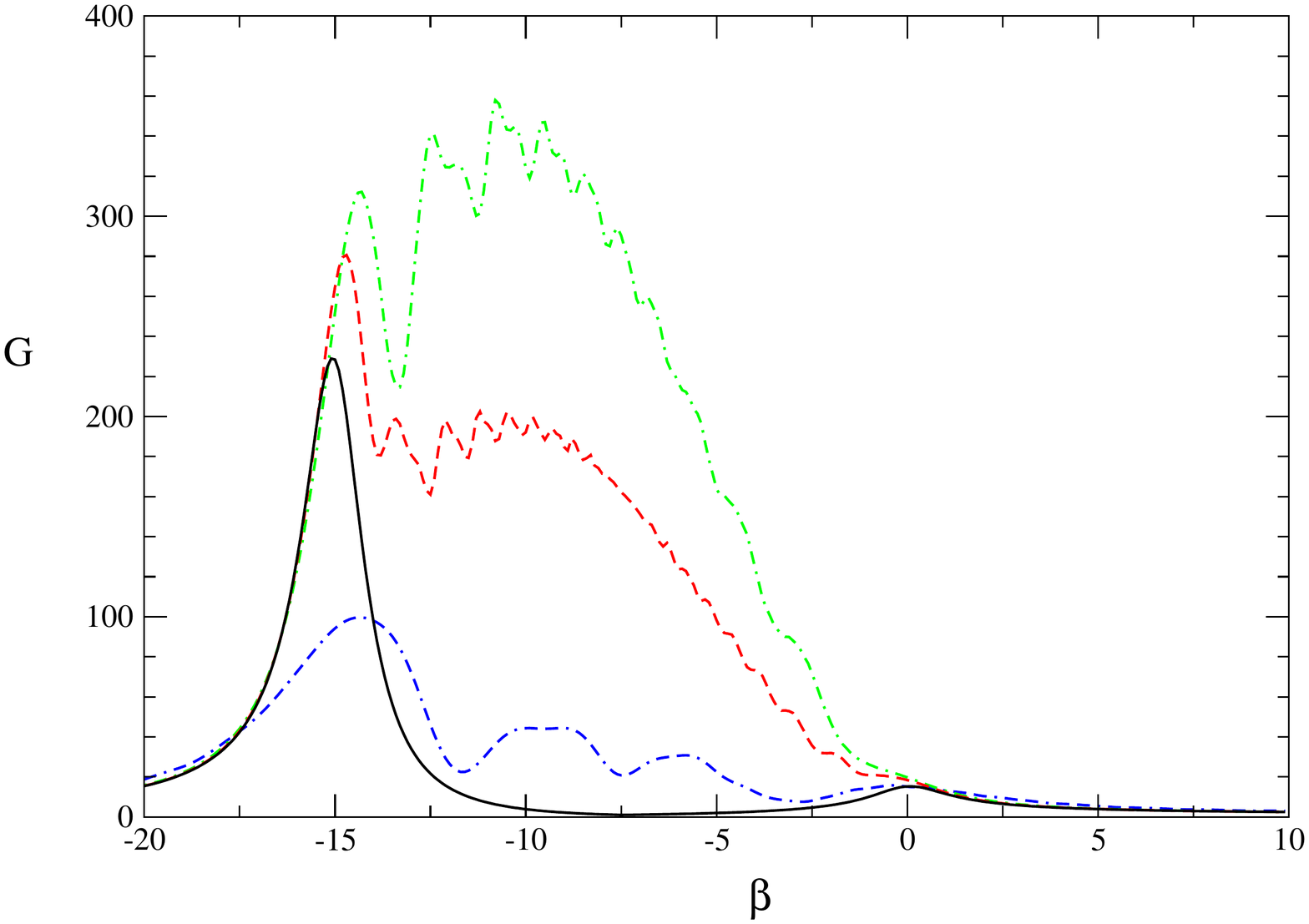}
\end{center}
\caption{
Optimal growth of perturbations in the shearing sheet approximation, $G(\beta)$, for
$\tau=10$, $q=3/2$. Perturbations are measured according to eq. (\ref{loc_norm}). 
Solid, dashed, dot-dashed and dot-dashed-dashed curves correspond to 
$R=0.1, 0.5, 1.0, 4.0$, respectively. Also, the solid curve virtually represents the
dependence given by eq.  (\ref{SFH_g}) for $\beta \lesssim -1$.
} \label{fig2}
\end{figure}

We start our research of optimal shearing harmonics in the particular case of 
Keplerian shear calculating $G(\beta)$ for different values of $R$. 
At first, in order to reveal basic features of non-modal growth only the acoustic energy is used to measure perturbations. 
Setting the optimisation time to the particular dynamical value $\tau=10$ 
we find that the non-modal growth is mostly an attribute of SFH with $\beta<0$ 
as would be expected from the theory of transiently growing vortices, see fig. (\ref{fig2}). 
Indeed, the solid curve obtained for small $R\ll 1$ virtually represents the incompressible 
growth factor given by eq. (\ref{SFH_g}) in the range of $\beta \lesssim -1$.
This result hints that performing the optimisation we just reproduce vortices in this case.
Note that in this range of parameters, $R\ll 1$ and $\beta \lesssim -1$,  
vortices and density waves exist separately from each other and the wave excitation is suppressed. 
Presumably, that is why the optimisation scheme 
approaches the pure vortex solution since any density wave can only return its energy to the flow
while being a leading spiral. The solid curve attains maximum at $\beta = -15 = -(3/2) \tau$, i.e.
for SFH that swings at $t  = \tau$.
At the same time, we notice that the optimal growth exceeds unity for $\beta \gtrsim 1$ as well. 
In the domain $\beta \gtrsim 1$ the two types of SFH are well distinguished again 
and we should suppose that the optimisation scheme approaches the density wave since
the vortex can only return its energy to the flow while being a trailing spiral.
However, the non-modal growth for $\beta\gtrsim 1$ is highly reduced in comparison with the case 
$\beta\lesssim -1$. We attribute this to the fact that the growth rate of density waves is 
proportional to $qt $ rather than to $(q t )^2$ as it is in the case of vortical SFH (see eq. (\ref{G_lowR})). 
Indeed, we can make use of the analytical results by \citet{Ch1997a} who derived how the acoustic energy
of density wave grows with time in the (non-rotating) shear, see their eq. (3.7a). 
In our notations it leads to the following dependence
\begin{equation}
\label{g_SD}
g = \left ( \frac{1 + (\beta + qt)^2}{1+\beta^2}\right )^{1/2}
\end{equation}
and it is assumed here that $R\ll 1$. Interestingly, it can be checked that eq. (\ref{g_SD}) perfectly recovers
the solid curve in fig.(\ref{fig2}) for $\beta>0$. 
For a sufficiently long time, $t  \gg \beta/q$, eq. (\ref{g_SD}) yields $g\approx qt / (1+\beta^2)$ 
affirming that $G\propto q t $ (not $\propto (q t )^2$ as for the maximum of $G(\beta)$ in the domain 
of negative $\beta$) and decreases monotonically as proceeding to large $\beta$.   
  
\begin{figure}
\begin{center}
\includegraphics[width=8.5cm,angle=0]{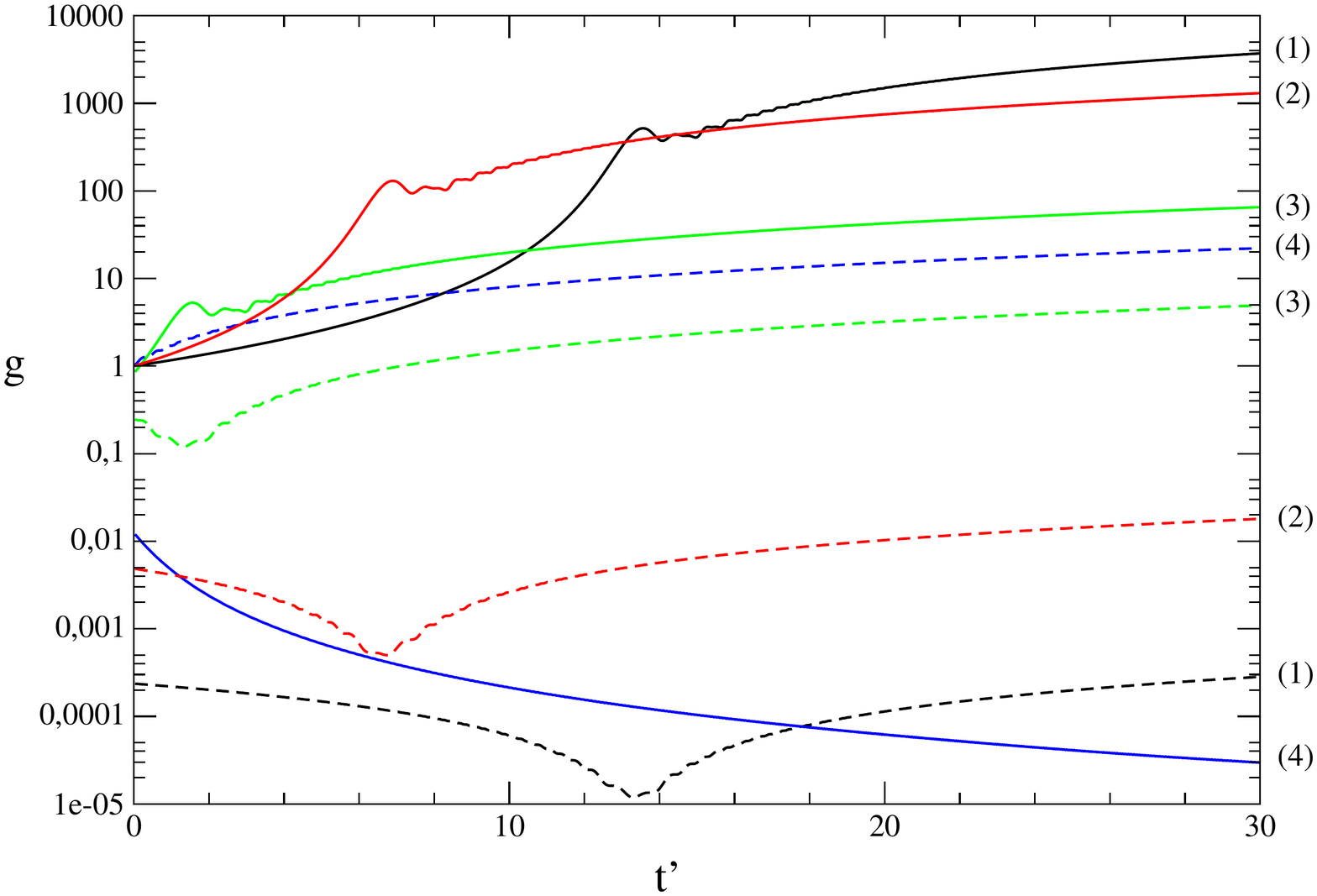}
\end{center}
\caption{Optimal perturbations in the shearing sheet approximation 
decomposed into vortical (solid curves) and wave (dashed curves) parts. 
Curves represent their growth factors, $g(t )$, and combined into pairs marked as (1), (2), (3), (4) 
corresponding to $\beta=-20, -10, -2, 2$, respectively. 
Perturbations are measured according to eq. (\ref{loc_norm}). 
The other parameters are $\tau=10$, $q=3/2$, $R=0.5$ (see dashed curve in fig.(\ref{fig2})).} 
\label{fig3}
\end{figure}

Now, turning to $R\lesssim 1$ we find that the maximum of $G(\beta)$ takes a greater value and 
slightly shifts towards SFH that swings a little before the optimisation instant. Moreover, 
a significant hump in $G(\beta)$ appears for $-15<\beta<0$.  
We suspect that this happens due to the phenomenon of wave excitation by vortex which arises for $R\sim 1$. 
This is what can enhance growth factor of vortical SFH right after it swings from leading to trailing
spiral since the emerged density wave extends the non-modal growth to trailing spiral phase. 
However, before we check our suspicion we would like to reveal the physical nature of 
optimal solutions using a rigorous algorithm.  
For that, we decompose the initial optimal SFH onto the vortex and the density wave. 
First, the potential vorticity perturbation of optimal SFH, $I$, is determined using eq. (\ref{vorticity}). 
Using the iterative method proposed in the Appendix by B05 we obtain values of $\hat u_x, \hat u_y$ and $\hat W$
at $t =0$ proportional to $I$ that extract the vortex from the optimal solution. 
This method converges provided that at $t =0$ SFH is outside the swing interval, see eq. (\ref{coupled_interval}).
If so, the remainder is nothing but density wave with $I=0$ being a solution of homogeneous part 
of eqs. (22) and (23) by HP. 
The result of decomposition is presented in fig.(\ref{fig3}) where we take several values of $\beta$ and 
corresponding optimal solutions lying on the dashed curve in fig.(\ref{fig2}). 
It is shown, that optimal solution is in fact a mixture of vortex and density wave.
Note that we have checked numerically that accuracy of method, determining the vortex and accuracy of optimisation
loop always lies well under the smallest of the amplitudes onto which the optimal solution is decomposed. 
As can be seen in fig.(\ref{fig3}), both vortex and wave constituting each pair swing simultaneously 
at the time defined by the value of $\beta$. As it should be, the leading spiral density wave changes from decay 
to growth as opposed to the vortex. While $\beta \ll -1$ the optimal SFH is almost a vortex rather than a wave.  
At the same time, when one sets $\beta>0$ there is no swing and 
the trailing spiral vortex decays from the very beginning as opposed to the density wave which becomes a cardinal 
component of optimal solution. Further, if an absolute value of $\beta$ becomes smaller 
the contribution of the secondary component increases.  
Thus, the overall conclusion can be drawn that in the domain where the non-modal growth attains its highest
rates (i.e. for $\beta \ll -1$) the initial optimal SFH is virtually indistinguishable from vortex 
carrying the same potential vorticity.

Armed with this knowledge we pay attention to solid curves of pair (1) and pair (2) in fig.(\ref{fig3}).
As was pointed out, these growth factors, $g(t )$, correspond to pure vortex solution at $t =0$. 
It becomes clear that since these vortices inevitably excite density waves at the instant of swing, these
density waves are the reason for an additional hump in the dashed curve in fig.(\ref{fig2}) in comparison with solid curve
in fig.(\ref{fig2}). As can be seen in fig.(\ref{fig3}), for $R=0.5$ the amplitude of the excited density wave 
is so large that $g$ grows almost monotonically all the time except a small peak at the time of swing, 
whereas the small scale vortex with $\lambda_y\ll H$ decays after it swings. 
That is why the difference in $G(\beta)$ between the cases $R=0.1$ and $R=0.5$ is so large in the 
range $-15<\beta<0$. However, the raise of optimal growth is also noticeable for $\beta \approx -15$ corresponding 
to swing right at the optimisation instant. The magnitude of this raise directly characterises
the additional amplitude of the density wave right after it was excited. As we have already mentioned, 
this amplitude is defined by the parameter $\epsilon$, see eq. (\ref{epsilon}), and attains maximum 
for $R\sim 1$. This result has been obtained analytically by HP. 
For $R=1$ (dashed-dotted curve in fig.(\ref{fig2})) the amplitude and the growth rate of the excited density wave 
becomes so high that the optimal SFH swinging approximately 1.5 times earlier than the optimisation instant
attains the highest optimal growth throughout the range of $\beta$. The corresponding value of $G(\beta)$ 
by 1.5 times exceeds its value in the small wavelength limit, $R\ll 1$.
At last, we plot $G(\beta)$ for the case $R=4$ representing a longer wavelength behaviour. We find that 
maximum of $G(\beta)$ returns to the position corresponding to SFH swinging at the optimisation instant 
what apparently conforms with the fact that the amplitude of the emerged density wave strongly decreases 
as $\epsilon$ becomes less that unity again. At the same time, the growth factor of vortex itself also decreases as
it is expected according to eq. (\ref{G_highR}) as function of $R$. 

\begin{figure}
\begin{minipage}[h]{0.96\linewidth}
\includegraphics[width=0.96\linewidth]{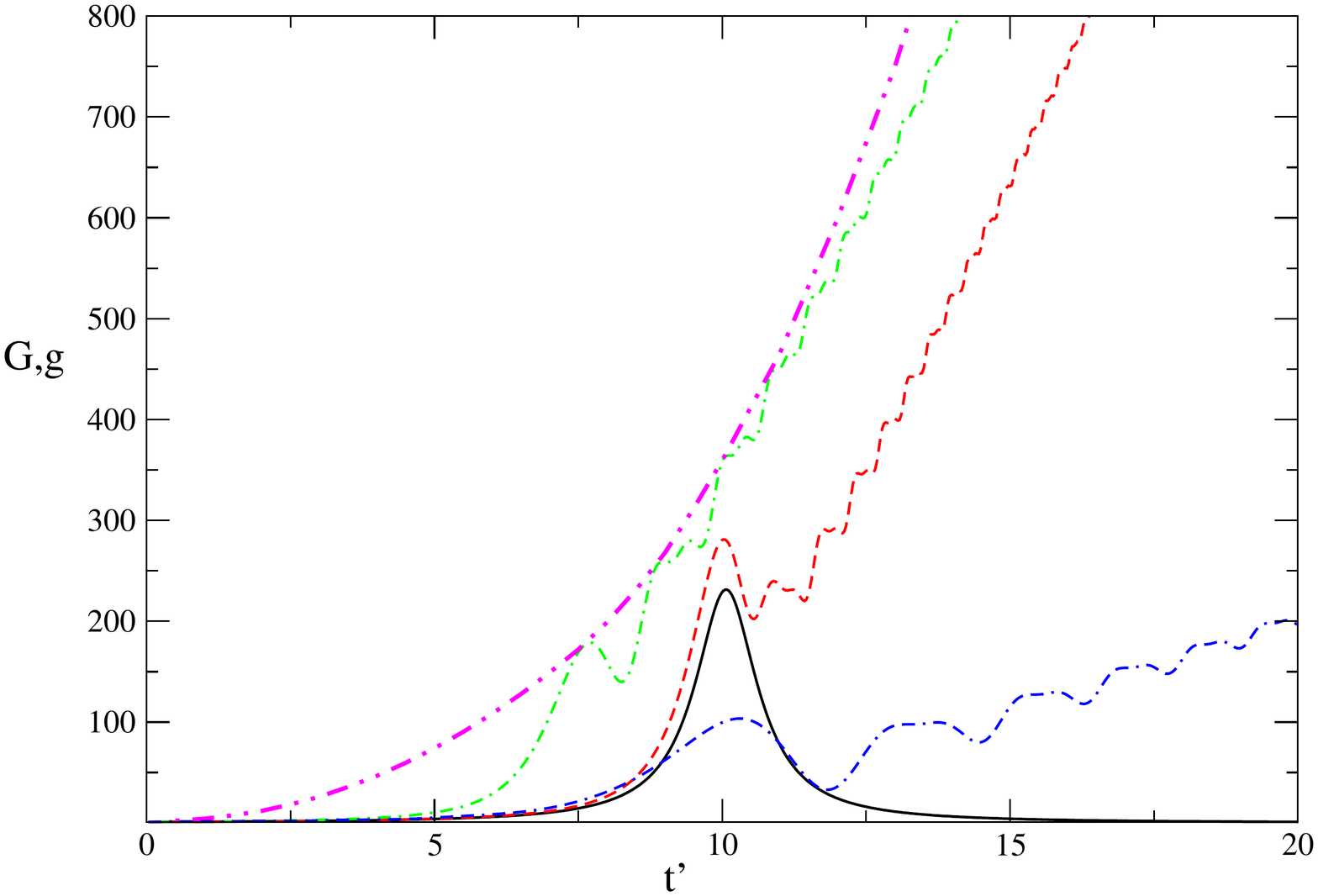} 
\end{minipage}
\vspace{1cm}
\vfill
\begin{minipage}[h]{0.96\linewidth}
\includegraphics[width=0.96\linewidth]{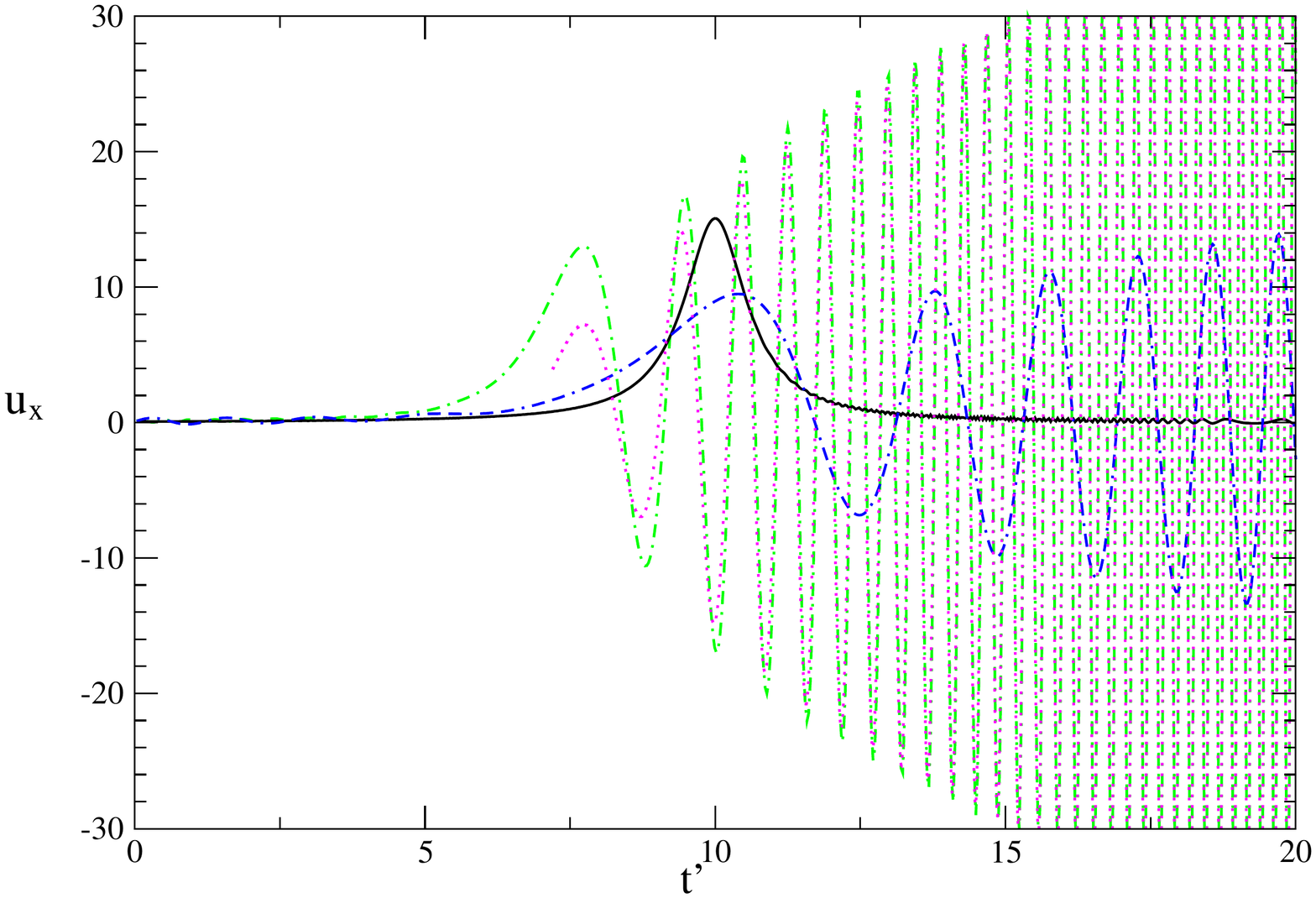} 
\end{minipage}
\vspace{1cm}
\vfill
\begin{minipage}[h]{0.96\linewidth}
\includegraphics[width=0.96\linewidth]{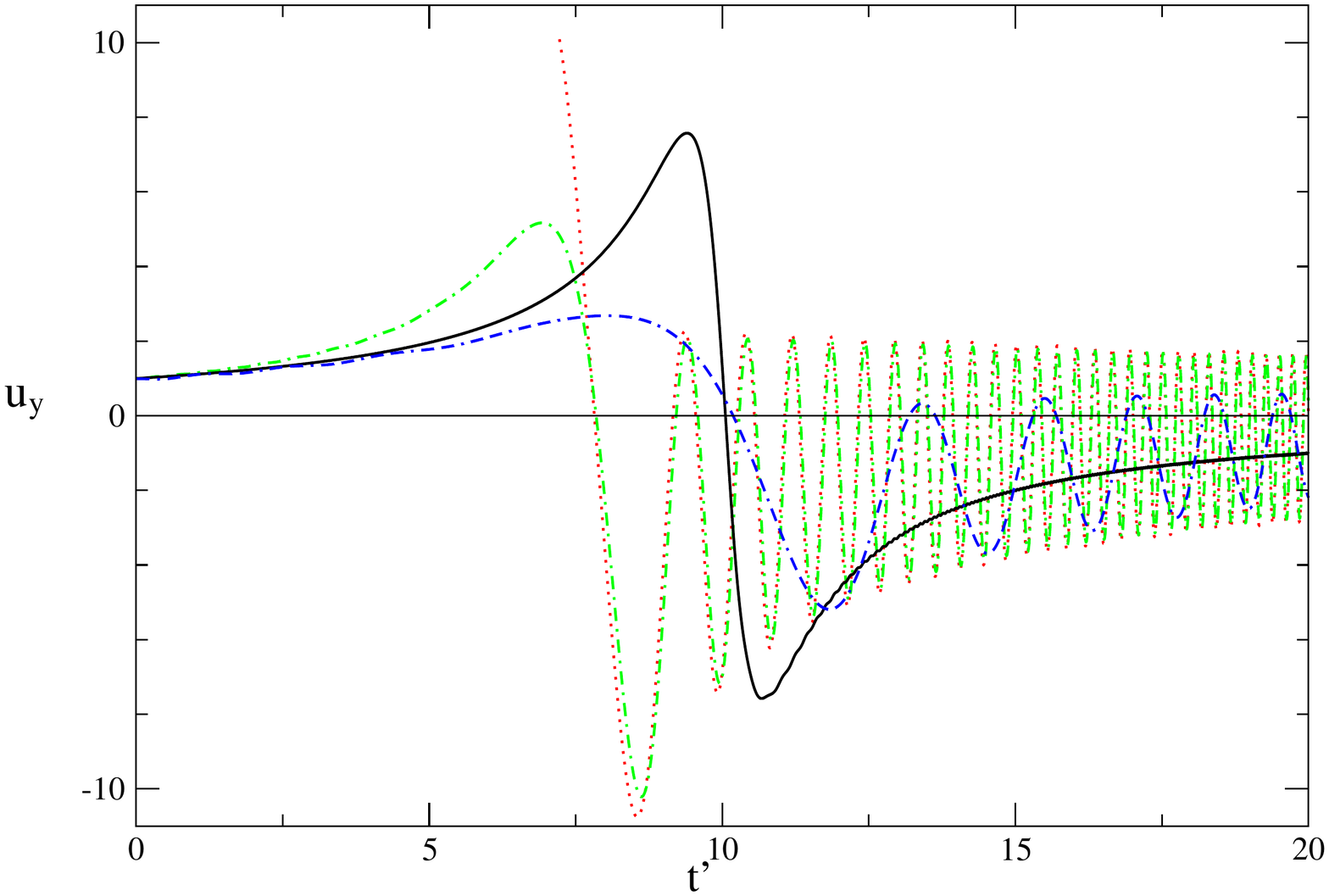}
\end{minipage}
\caption{Top panel: Solid, dashed, dot-dashed and dot-dashed-dashed curves are the growth factors of optimal perturbations
determined for $R=0.1, 0.5, 1.0, 4.0$, respectively. Each optimal perturbation is taken for value of $\beta$ 
which corresponds to  $G\equiv\max\{G(\beta)\}$, look at fig.(\ref{fig2}). Also, the solid curve virtually represents the
dependence given by eq.  (\ref{SFH_g}) for fixed $\beta=-15$.
Dot-dot-dashed curve is the maximum of $G$ over all $R$ as a function of time, see 
dot-dot-dashed curve in fig. (\ref{fig5}). 
Middle and bottom panels: radial and azimuthal velocity perturbations depending on time. Solid, dot-dashed and 
dot-dashed-dashed curves correspond to $R=0.1, 1.0, 4.0$, respectively. Also, dotted curve represents 
the sum of analytical solutions for vortex and wave after the instant of swing 
which are borrowed from \citet{HP09_1}, look at their eqs. (31), (52) 
and (57), and produced with the same $I$, $k_x$ and $k_y$ as for optimal $u_x$ and $u_y$. 
Perturbations are considered in the shearing sheet and measured according to eq. (\ref{loc_norm}). 
For all curves excluding the dot-dot-dashed one in the top panel the other parameters are $\tau=10$, $q=3/2$.
} 
\label{fig4}
\end{figure}

In order to illustrate how the evolution of growth factors of optimal SFH changes with $R$ we plot $g(t)$ for 
the same values of $R$ as in fig. (\ref{fig2}) taking $\beta=\beta_{max}$ corresponding to maxima of curves of
$G(\beta)$ in fig. (\ref{fig2}). Since $\beta_{max} \ll -1$ for all curves, $g(t )$ virtually 
represents the evolution of corresponding vortices with the same potential vorticity perturbation, see 
top panel in fig. (\ref{fig4}). 
As soon as we take $R$ close to unity, they excite density waves with the acoustic energy growing linearly 
with $t $ (cf. eq. (\ref{g_SD})) at long times. The dot-dashed curve in fig.(\ref{fig4}) demonstrates 
the case $R=1$ which, in fact, gives the highest possible growth over all $R$ for fixed optimisation time, 
see fig. (\ref{fig6}) in the next section. Moreover, the optimal solution for $R=1$ is perfectly 
recovered by the analytical solution by HP in spite of the fact that $R=1$ and, consequently, 
$\epsilon \sim 1$ for the Keplerian flow corresponds to the worst case for the asymptotical analytics employed by HP. 
Indeed, this is shown in the middle and the bottom panels in fig. (\ref{fig4}) where we plot 
the sum of eq. (\ref{vort_u_x}) and eq. (57b) by HP standing for $\hat u_x(t )$ and 
the sum of eq. (\ref{vort_u_y}) and eq. (52) by HP standing for $\hat u_y(t )$ (dotted curves). 
Note that these expressions are evaluated for the potential vorticity perturbation of optimal SFH. They are 
plotted for $t $ longer than the optimisation time since before the optimal SFH swings from leading to 
trailing spiral it is described by nearly vortical solution.

\begin{figure}
\begin{center}
\includegraphics[width=8.5cm,angle=0]{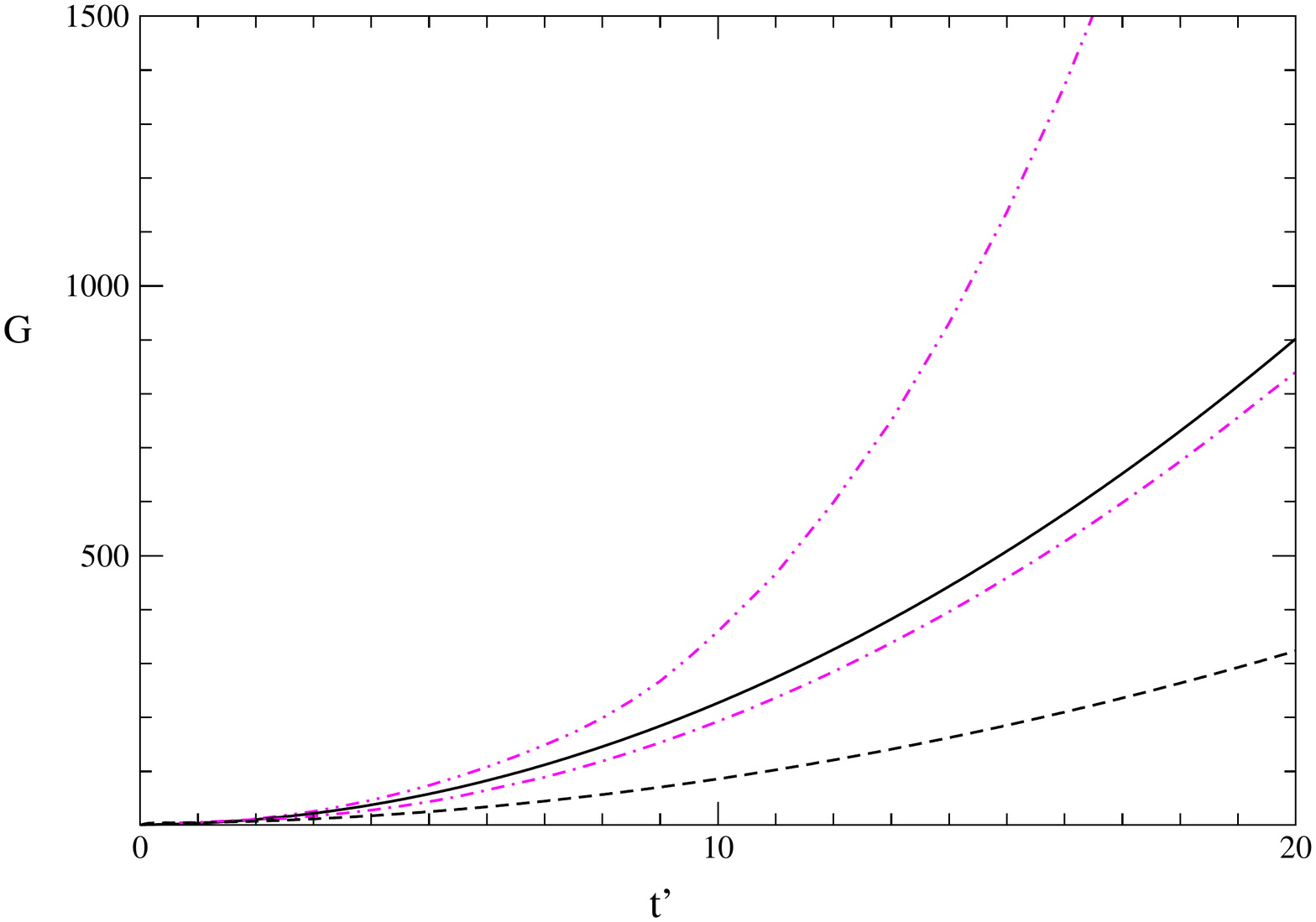}
\end{center}
\caption{Solid and dashed curves show the optimal growth as a function of time taken in the limit of $R\ll 1$ 
for local perturbations measured according to eq. (\ref{loc_norm}) and eq. (\ref{loc_norm_2}), respectively. 
Dot-dot-dashed and dot-dashed curves show the maximum of optimal growth over all $\beta$ and all $R$ 
as a function of time for perturbations measured according 
to eq. (\ref{loc_norm}) and eq. (\ref{loc_norm_2}), respectively.
The shear rate is $q=3/2$.} 
\label{fig5}
\end{figure}

Back to the top panel in fig. (\ref{fig4}), we find that $g(t )$ for $R=1$ is tangent to the dashed-dot-dot curve
in the vicinity of the optimisation instant. The dashed-dot-dot curve is defined as an optimal growth, $G(t )$, 
maximised over all $R$ (i.e. over all SFH with the specific $k_x$ and $k_y$), see also fig. (\ref{fig5}). 
None of initial local perturbations can grow in the acoustic energy higher than it is limited by this curve. 
In the fig. (\ref{fig5}) this general bound of transient growth in compressible Keplerian flow is plotted 
for both norms of perturbations we have defined in this work. There are also two additional curves that 
represent the optimal growth in the limit of small scale perturbations, $\lambda_y \ll H$, measured
either by eq. (\ref{ac_en}) or by eq. (\ref{ac_en2}) as well. Let us remember that the case $\lambda_y \ll H$ is 
equivalent to transient growth of local perturbations with divergence-free velocity field, see Appendix B for details.
It turns out that independently of the choice of norm the compressible perturbations are able to 
grow significantly faster (by a factor increasing with timespan) than incompressible perturbations.
Again, independently of the norm choice perturbations with a characteristic length scale or order of the 
disc thickness ($\lambda_y\sim H$) grow most rapidly at the fixed time interval what takes place due 
to the ability of vortices with $\lambda_y\sim H$ to excite the strongest density waves.
Thus, the compressibility of medium is factor that cannot be ignored while studying the 
transient growth phenomenon in astrophysical discs.  
Here it should be mentioned that simulations performed by \citet{HP09_2} in the shearing sheet model 
confirmed the emergence of density waves generated by vortices having the turbulent origin. 
They also verified that this phenomenon becomes most prominent for the wavelengths $\sim H$, see 
fig. (7) of their paper. 
In this study we come to similar conclusions looking for an extreme solutions to the initial value problem 
for linear perturbations. However, the non-linear evolution of the excited density wave extracted 
by \citet{HP09_2} shows a significant damping contrary to the linear solution. The latter indicates
that the solutions represented by dot-dashed curve in fig. (\ref{fig4}) must be saturated stronger than all others 
by the non-linear effects in a well-developed turbulence.

Despite the argument that the scale of order of the disc thickness has been revealed to be the 'optimal' in context of
non-modal growth in Keplerian flow we believe it is quite important to consider in a more detail 
the case of large scale perturbations, $R\gg 1$, when the density wave excitation is suppressed just as in 
the 'incompressible' case, $R\ll 1$. 
Foremost, this can be justified by the fact that
the large scale perturbations are not a subject to fast dissipation in realistic viscous (turbulent) shear flow.
Our estimate shows that dissipation time $\propto \alpha^{-1/3} R^{2/3}$, see eq. (\ref{T_max}), and consequently
the highest possible growth for all timespans, $G_{max}$, does not depend so sharply on $R$ as it happens to 
$G$ for the fixed timespan in the inviscid fluid, cf. eq. (\ref{G_highRt}). Moreover, 
as we have already assessed at the very end of section 3.2.2, for realistic thin accretion discs with an aspect ratio 
$0.01 \lesssim \delta \lesssim 0.1$ with insufficient turbulent viscosity, $\alpha \lesssim \delta$, $G_{max} > 1$  
for all possible values of $\lambda_y$ up to $\lambda_y\sim r$, where $r$ is the radial scale of disc. 
Secondly, we have shown above that the most rapidly growing (i.e. optimal) perturbations are nearly the vortices, 
thus, any preliminary chaotic motions in disc acquiring the large scale potential vorticity perturbation become
natural seeds for large scale (compressible) vortices. Through the linear transient growth these vortices
can provide an extra angular momentum transfer in weakly turbulent discs. 

Additionally, we would like to mention that recent MHD simulations of turbulence in accretion discs 
performed on intermediate and global spatial scales show that a significant fraction of the accretion stress
is contained in azimuthal modes with $k_y H \lesssim 1$, see for example 
the plots with autocorrelation functions of Maxwell and Reynolds stresses in \citet{Armitage2012} and  
the plot with the toroidal power spectrum of the Maxwell stress as well as the plot with the
total accretion stress on small azimuthal scales relatively to the same quantity on large azimuthal scales in 
\citet{Armitage2011}. Whether it is unclear, to what degree the large scale magnetic field 
is important in this situation, the transiently growing large scale vortices considered in this work (see the next section) 
may give an independent contribution to non-local transfer of disc angular momentum.
However, in order to confirm or discard this guess it could be worthwhile to perform a study similar to that of
\citet{HP09_2} who considered particular spatial Fourier amplitudes of perturbations extracted
from their (local) simulations of MRI turbulence. A similar procedure applied to the result of simulations
on intermediate and global spatial scale, $\lambda_y\gg H$, could reveal whether the large scale vortices 
are responsible for an additional accretion stress due to their transient growth as they swing from leading to trailing 
spirals. Another fairly simplified approach to this issue is to regard the nonlinear contribution of turbulence
as an external noise imposed in disc. The latter has been successfully employed by \citet{IK2001} in 
application to incompressible global perturbations in Keplerian disc. An extension of their study to 
perturbations with $\lambda_y\gg H$ in thin discs would be quite worthy.    

For these reasons, in the next section we would like to investigate the large scale vortex transient 
dynamics employing the shearing sheet approximation as well as the global treatment of optimisation 
problem formulated previously (see section 3.1). Let us emphasise that the optimisation scheme in global 
approach allows to determine a unique radial profile of optimal solution with specified azimuthal wavenumber only, 
see fig. (\ref{fig1}). At the same time, in the shearing sheet model the optimisation scheme
converges to a single SFH, i.e. to the solution with specified $k_x$.

\subsection{Extension to global spatial scale}

In this section, whenever we mention parameter $R$ in the context of global dynamics, we mean its following analogue 
\begin{equation}
\label{R_gl}
R=\frac{3/2}{m a_*|_{r=1} }
\end{equation}
which equals to local version of $R$ in the vicinity of $r=1$ in case of the Keplerian shear.
In order to make comparisons with the shearing sheet model, we employ the homogeneous disc with uniform
$\Sigma$ and aspect ratio, $\delta$, see section 3.1.2. This allows to study solely the influence of non-zero
background vorticity gradient discarded in the shearing sheet as well as the corrections due to the cylindrical geometry.
It is implied that local dynamics is considered in the vicinity of the inner boundary of disc, $r=1$,
and the time in local problem is measured in units of the inversed Keplerian frequency at $r=1$.

\begin{figure}
\begin{center}
\vspace{1cm}
\includegraphics[width=8.5cm,angle=0]{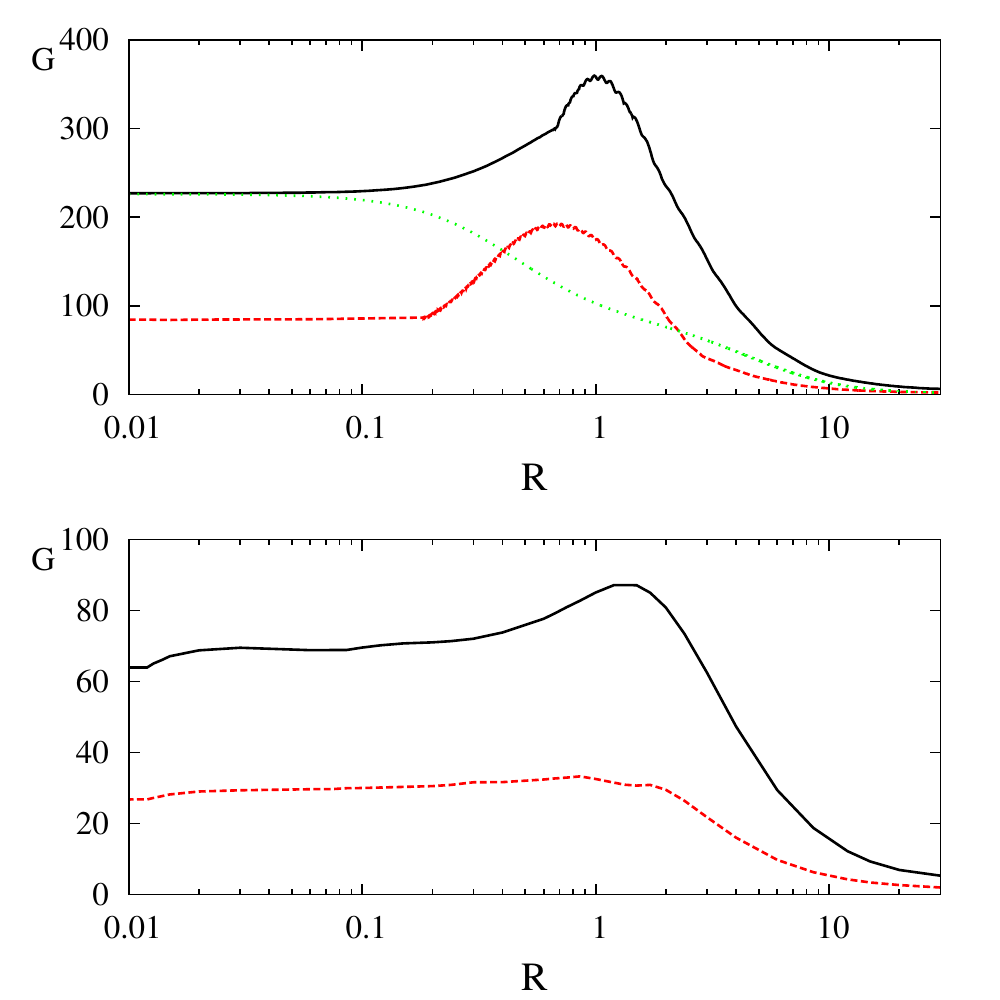}
\end{center}
\caption{The optimal growth, $G\equiv\max\{G(\beta)\}$, as a function of $R$. 
The top panel represents perturbations considered in the shearing sheet model with $q=3/2$ and measured 
according to eq. (\ref{loc_norm}) (solid curve) and eq. (\ref{loc_norm_2}) (dashed curve), whereas 
the bottom panel represents perturbations considered globally with $m=5$ in homogeneous Keplerian disc with $n=3/2$
and measured according to eq. (\ref{ac_en}) (solid curve) and eq. (\ref{ac_en2}) (dashed curve).
Dotted curve in the top panel reproduces the analytical dependence given by eq. (\ref{G_compr}). 
The optimisation time is $\tau=10$.}
\label{fig6}
\end{figure}

First, in the top panel of fig. (\ref{fig6}) we show the slice of $G(R)$ obtained in the shearing sheet model 
of the Keplerian flow for fixed optimisation time, $\tau=10$. 
Clearly, the maxima of curves of $G(R)$ representing two different norms 
of perturbations lie on the dot-dot-dashed and dot-dashed curves in fig. (\ref{fig5}).
In case of norm (\ref{ac_en}) the maximum of $G$ is attained close to $R=1$ 
whereas in case of norm (\ref{ac_en2}) it is shifted to a smaller value of $R$.
As it has been discussed in the previous section, the humps that we find in $G(R)$ for both norms of perturbations
emerge due to excitation of density waves by vortices which are the major components of initial optimal SFH. 
Also, we find the breaks on each curve of $G(R)$. 
In the case of norm (\ref{ac_en}) the break is located approximately at $R\approx 0.7$ whereas
in the case of norm (\ref{ac_en2}) it is shifted to $R\approx 0.2$. 
These breaks are caused by a 'jump' from local maximum of $G(\beta)$ corresponding to SFH swinging closely 
to the optimisation instant to another local maximum of $G(\beta)$ corresponding to SFH swinging earlier
in time. It happens when the density wave excited by the latter SFH is so high 
that the latter maximum becomes larger than the former one, refer back to fig. (\ref{fig2}). 
In the top panel of fig. (\ref{fig6}) we also plot an additional curve that reproduces our estimate of $G$ 
given by eq. (\ref{G_compr}) which represent solely the vortices. 
This curve makes distinctive the effect of density waves emergence for perturbations with $\lambda_y \sim H$.
Note that for $R\gg 1$ the analytical results for vortices yields a moderate underestimation of $G$ despite 
the fact that density wave excitation is already exponentially suppressed. This happens because the swing 
interval becomes too wide and analytical solution for vortices (see eq. (\ref{vort_u_x}-\ref{vort_W}))
diverges with precise numerical solution, see also fig. (\ref{fig8}).

\begin{figure}
\begin{center}
\vspace{1cm}
\includegraphics[width=8.5cm,angle=0]{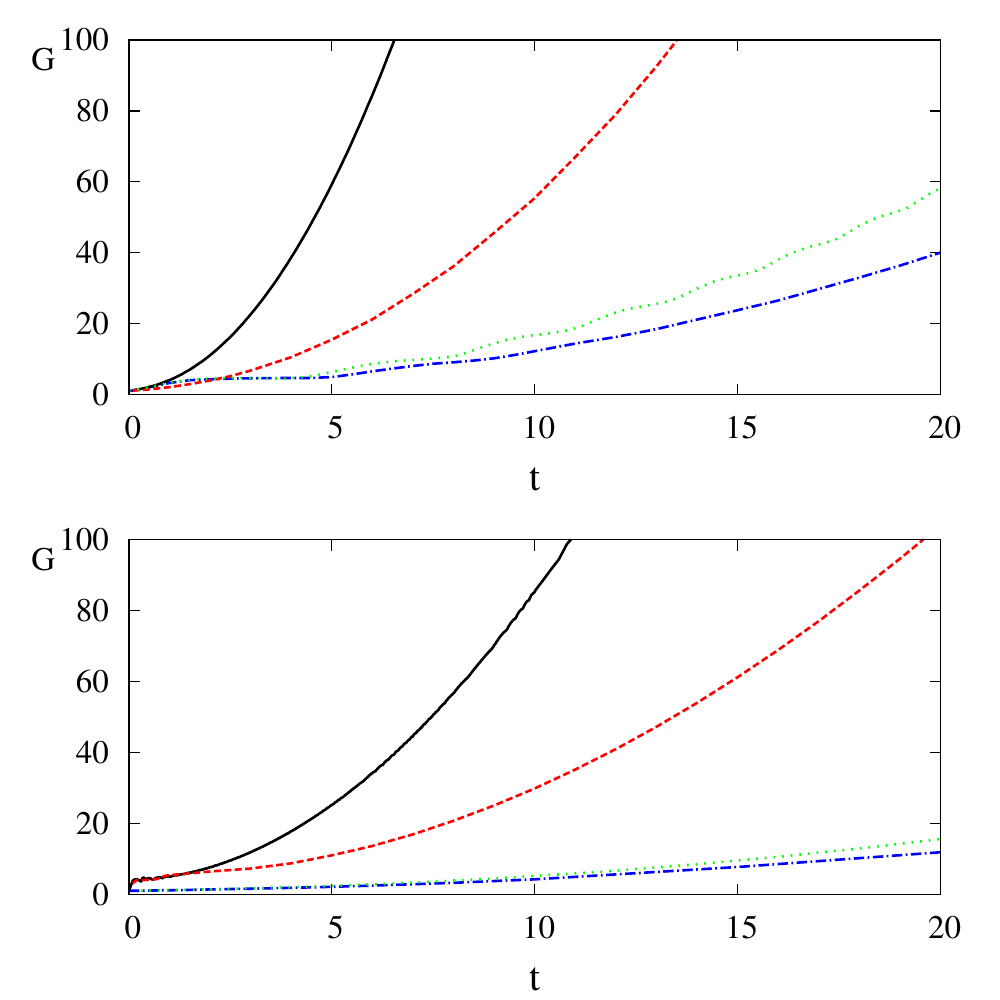}
\end{center}
\caption{Comparison of optimal growth for local and global perturbations. Top panel 
displays perturbations measured according to eqs. (\ref{loc_norm}) and 
(\ref{ac_en}) in local and global cases, respectively, whereas 
in the bottom panel they are measured by eqs. (\ref{loc_norm_2}) and 
(\ref{ac_en2}) instead. Solid and dotted curves in both panels show $G\equiv \max\{G(\beta)\}$
obtained in the shearing sheet model as a function of time for specified $R=0.12$ and $R=12$, respectively.
Dashed and dashed-dotted curves in both panels show the optimal growth obtained globally in 
case $m=5$ for the same $R=0.12$ and $R=12$, respectively.   
In the local case it is assumed that $q=3/2$. Accordingly, the global perturbations 
are considered in the Keplerian homogeneous disc with $n=3/2$.
} \label{fig7}
\end{figure}

In the bottom panel of fig. (\ref{fig6}) the dependencies $G(R)$ obtained in the global problem are shown. 
Setting a particular value of (global) azimuthal wavenumber, $m=5$, and a particular polytropic index, $n=3/2$, 
we alter $R$ varying the disc aspect ratio, $\delta$. 
For example, in order to get the optimal solution for $R=10$ we must set $\delta=0.06$ in dynamical equations, etc.
Corversely, $R<1$ formally corresponds to a thick disk with $\delta>1$.
Basically, in global configuration the transient growth of perturbations with $R<1$ is considerably suppressed in 
comparison with the local case. Qualitatively,  the curves of $G(R)$ in global and local approach resemble each other 
except that for the second norm, eq. (\ref{ac_en2}), the hump produced by density waves excitation almost vanishes. 
Also, we see that second norm always yields a smaller optimal growth in comparison with perturbations
measured by the acoustic energy (see figs. (\ref{fig5}) and (\ref{fig7}) as well).
Despite that it is found that $G\gg 1$ in all possible variants.
On the other hand, we find that in the opposite case, $R\gg 1$, the difference between global and local optimal 
growth is noticeably reduced. In particular, if perturbations are measured by the acoustic energy
the optimal growth evaluated at $R=0.1$ is larger than those evaluated at $R=10$ roughly by a factor of $10$ in 
the shearing sheet model, see solid curve in fig. (\ref{fig6}). But the same difference for global perturbations 
is given roughly by a factor of $3.5$.  
This property of global transient growth it better to check 
looking how $G$ depends on time for particular values of $R$.
We plot $G(t)$ in fig. (\ref{fig7}) using both local and global optimisation method which is again employed 
for two choices of norm of perturbations (top and bottom panels). Indeed, the curves obtained for large $R=12$ 
settle much closer to each other than those obtained for small $R=0.12$ in both panels of the figure. 
The optimal growth in the marginal case $m=1$ is not plotted in fig. (\ref{fig7}) 
but, as we have checked, for small $R=0.12$ $G(t=20)\approx 30$, whereas for large $R=12$ 
$G(t=20)\approx 20$ what is only $1.5$ times smaller. Actually, this implies that the global 
large scale vortices, $R\gg 1$, we consider in this work exhibit the transient growth almost 
comparable to those we know since paper by \citet{IK2001} who presented $G(t)$ for incompressible 
global perturbations in Keplerian flow, see their fig. (1).
But as we have found previously, this is not the case in local approach: small scale optimal SFH 
($R\ll 1$) grows much more rapidly than large scale optimal SFH ($R\gg 1$), 
see the results of section 4.1 and top panel of fig. (\ref{fig6}).
Since we have mentioned the calculations by \citet{IK2001}, 
one should bear in mind, that we present the dynamics with quite short 
optimisation timespans $\tau<20$ which corresponds to less than $\sim 3$ Keplerian orbits at the inner boundary of disc. 
For comparison, \citet{IK2001} measured time in Keplerian periods at $r=10r_0$ 
which corresponds to rescale of dimensionless time by a factor of $\sim 180$ if changing from their to our units. 
Thus, we would actually see the comparable magnitudes of transient growth of our large scale global vortices with
$R\gg 1$ if corresponding $G(t)$ were plotted in their fig. (1).

Additionally, we test our global numerical scheme checking that in the limit of large $m\gg 1$ (freezing the value of $R$) 
and small size of computational domain (nearby $r=1$) it reproduces the solid curves in both panels of fig. (\ref{fig7}), 
i.e. we check that dashed curves approach solid curves in the limit $m\gg 1$ (and fixed $R$). 
Furthermore, the optimisation of global incompressible perturbations described 
in the Appendix B is employed to make one more independent check of our basic numerical scheme. 
We make sure that in case of $m=5$ the iterative loop based on ${\bf A}$ and ${\bf A}^\dag$ given by eqs. 
(\ref{A_expl_vort}) and (\ref{A_adj_expl_vort}), respectively, yields $G(t)$ that virtually recovers 
the dashed curve in the top panel of fig. (\ref{fig7}).
Also note that solid curves in fig. (\ref{fig5}) and in top panel of fig. (\ref{fig7}) virtually recover 
the analytical $G(t)$ given by eq. (\ref{G_lowR}). 

Getting back to the global large scale perturbations, $R\gg 1$, it can be noticed that 
for short time intervals less than one Keplerian period at the inner boundary of disc ($t\lesssim 5$)
the optimal growth of perturbations measured by the acoustic energy 
acquires a flat segment corresponding to $G\approx 4$ in both local and global
problems, see dotted and dashed-dotted curves in the top panel of fig. (\ref{fig7}).
On the contrary, in case of $R\ll 1$, $G\to 1$ self-similarly while $t\to 0$, see dashed curve in the same panel. 
This discordance would become particularly distinctive if one plotted the marginal case of
$m=1$ with $a_* \to \infty$ standing for the Fourier global mode with the least azimuthal wavenumber 
in incompressible fluid contrary to the marginal case of $m=0$ with finite $a_*<\infty$
standing for compressible axisymmetric perturbation ($R\to \infty$). In the former situation 
we would get $G<4$ up to $t\simeq 5$, 
whereas, in the latter situation we would get $G=4$ of all time intervals $t \gtrsim 1$.
The point is that in contrast to solenoidal planar perturbations compressibility allows for existence of one-dimensional radial motions 
in the perturbed flow. Moreover, they have an optimal configurations, which are able to grow by 
a factor of $\sim 4$ in Keplerian disc, as measured by their total acoustic energy. 
This factor is nothing but the squared ratio of epicyclic frequency for rigid rotation 
to epicyclic frequency of shear flow under consideration. 
In the Appendix A we interpret this fact and give a detailed description of 
optimal axisymmetric perturbations. 
We argue that at short time intervals the epicyclic motions in the rotating shear flow (see Appendix A)
are responsible for the flat segment that emerges on curves of $G(t)$ as $R$ goes to infinity. 
At the same time, in the case when the compressible perturbations are optimised according to eq. (\ref{ac_en2})
(bottom panel in fig. (\ref{fig7})), the curves of optimal growth have similar shape as 
those for perturbations in incompressible fluid. 
Thus, as $R$ increases $G\to 1$ for all time intervals. 
This is an expected result since the norm (\ref{ac_en2})
has been introduced to exclude the non-modal behaviour of axisymmetric perturbations (see sect. 3.1.1 of this paper).

\begin{figure}
\begin{center}
\includegraphics[width=8.5cm,angle=0]{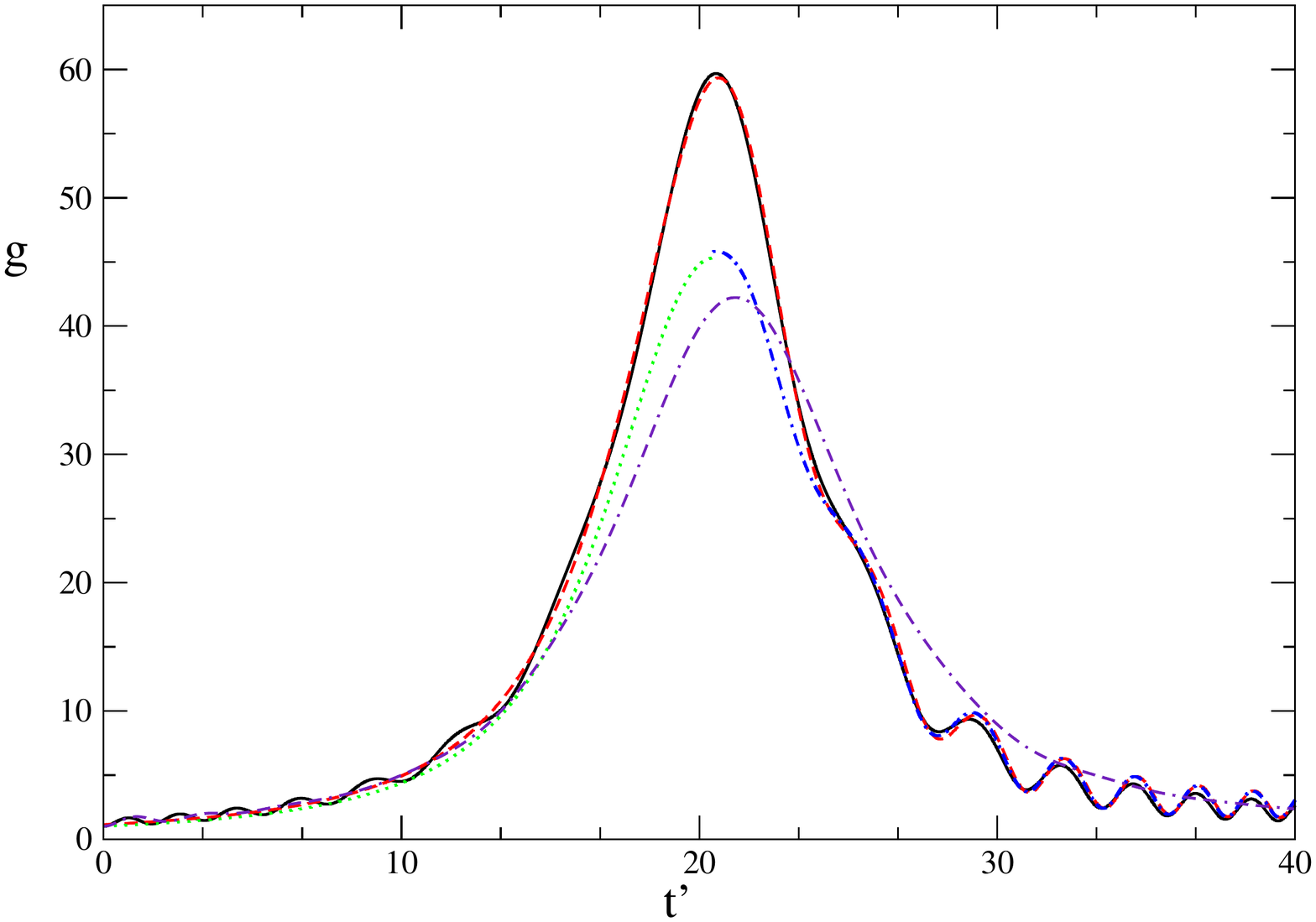}
\end{center}
\caption{Growth factors of various kinds of large scale perturbations $\lambda_y\gg H$ measured according to 
eqs. (\ref{ac_en}) and (\ref{loc_norm}). 
Solid curve stands for optimal perturbation obtained in the shearing sheet model with 
parameters $R=12, \tau=20, q=3/2$. Dashed curve stands for perturbation advanced numerically 
using the vortex initial conditions, look Appendix by B05. Dotted and dashed-dotted curves represent 
the analytical solutions for vortex before the instant of swing and for sum of vortex and density wave after 
the instant of swing, look at eqs. (31), (52) and (57) by HP. 
Finally, dot-dashed-dashed curve represents global optimal perturbation
with $m=5$ and $R=12$ obtained in homogeneous Keplerian disc with $n=3/2$.} 
\label{fig8}
\end{figure}

To conclude this section we focus once again on the large scale vortices, $R\gg 1$.
Our incentive is to verify the validity of their analytics outlined in section 3.2.2.
In fig. (\ref{fig8}) the growth factors, $g(t)$, obtained in various approaches are plotted. 
We choose a particular value $R=12$ for the shearing sheet model. 
We set $m=5$ for global calculations what means that one takes $\delta=0.05$ for the background since
the same value of the global analogue of $R$ is implied. The optimisation timespan is $\tau=20$.
First, the optimal SFH is represented
by solid curve. It is obtained by means of the optimisation procedure in the shearing sheet approximation 
for $\beta\approx -30$ what corresponds to $\max\{ G(\beta)\}$ for fixed $R$. 
Since $|\beta|>R$ the iterative method described in the Appendix by B05 has a good convergence at the moment $t=0$
and we employ it to determine the initial conditions for pure vortex with the same potential vorticity
as for our optimal SFH (see also the description in section 4.1 and fig. (\ref{fig3})). 
Using these initial conditions we advance perturbations numerically and obtain the dashed curve in fig. (\ref{fig8}).
Clearly, one gets an excellent agreement with solid curve since for $\beta\ll -1$ the optimal solution
is almost a vortex. 
Further, in accordance with theory by HP we find that a weak signature of density wave
excitation appears well after the vortex decays giving its energy back to the flow, cf. the intensity of the 
excited density wave with the case $R=4$ plotted in fig. (\ref{fig4}), top panel. 
Then, we plot the analytical solution by HP employing the norm (\ref{loc_norm}) and constructing the result 
from two different curves. The first one corresponds
to the analytical solution for vortex, eq. (31) by HP or eqs. (\ref{vort_u_x}-\ref{vort_W}) in this work, obtained
for known $k_x$ and $k_y$. The second one corresponds to the sum of the analytical solution for vortex and 
the analytical solution for excited density wave, eq. (52-57) by HP. 
As one can see in the plot, this analytical (composite) curve recovers well the numerical solutions
everywhere except the zone around the instant of swing. 
The latter is expected since as we have discussed in section 3.2.1 the existence of vortices becomes
ill-defined inside the swing interval given by eq. (\ref{coupled_interval}).  
In the particular case displayed in fig. (\ref{fig8}) $t _{s_1}=12$ and $t _{s_2}=28$. 
However, the actual size of the interval where analytics diverges with the precise solution 
luckily appears to be at least two times less than it follows from eq. (\ref{coupled_interval}). 
That is why the underestimation of $G$ from analytics is not dramatic in spite of the fact that 
the condition (\ref{coupled_cond}) is not fulfilled and $R \sim |\beta|/2$. 
Another encouraging thing is that actual growth of vortex has been found to be somewhat larger 
than it follows from analytics. 
Moreover, the approximate expression for $G$ given by eq. (\ref{G_highRt}) which is valid 
in the limit of large $R\gg 1$ and long $t\gg 1$, see section 3.2.2, yields the value of $G\approx 57$ in our 
particular case which is intermediate between the analytical and numerical values. 
At last, we plot the result of global optimisation which exhibits the transient growth 
close to the analytical estimate\footnote{Also note that inertial-acoustic modes would have
almost constant $g\approx 1$ if were plotted here since their increments (decrements) are too small in thin Keplerian 
disc, see e.g. \citet{GN1985} or \citet{Kato1987}.}. 

As we discuss in the end of section 4.1, in disc with small viscosity, $\alpha<\delta$, 
the vortices $\lambda_y\gg H$ have the ability to exhibit the transient growth on all scales up to 
the highest one, $\lambda_y \sim H/\delta$. Additionally to analytical estimations of $G_{max}$ given 
in section 3.2.2 we check its magnitude using the numerical optimisation in the shearing sheet model. We
determine $G$ for the timespan (\ref{T_max}). Particularly, we find that the maximum value of $G_{max}$
in the whole range of $R>1$ attains $\sim 250$ and $\sim 1000$ for $\alpha=10^{-3}$ and $\alpha=10^{-4}$, respectively.

\subsubsection{Particular case of formally inviscid Shakura-Sunyaev accretion disc}

\begin{figure}
\begin{center}
\vspace{1cm}
\includegraphics[width=8.5cm,angle=0]{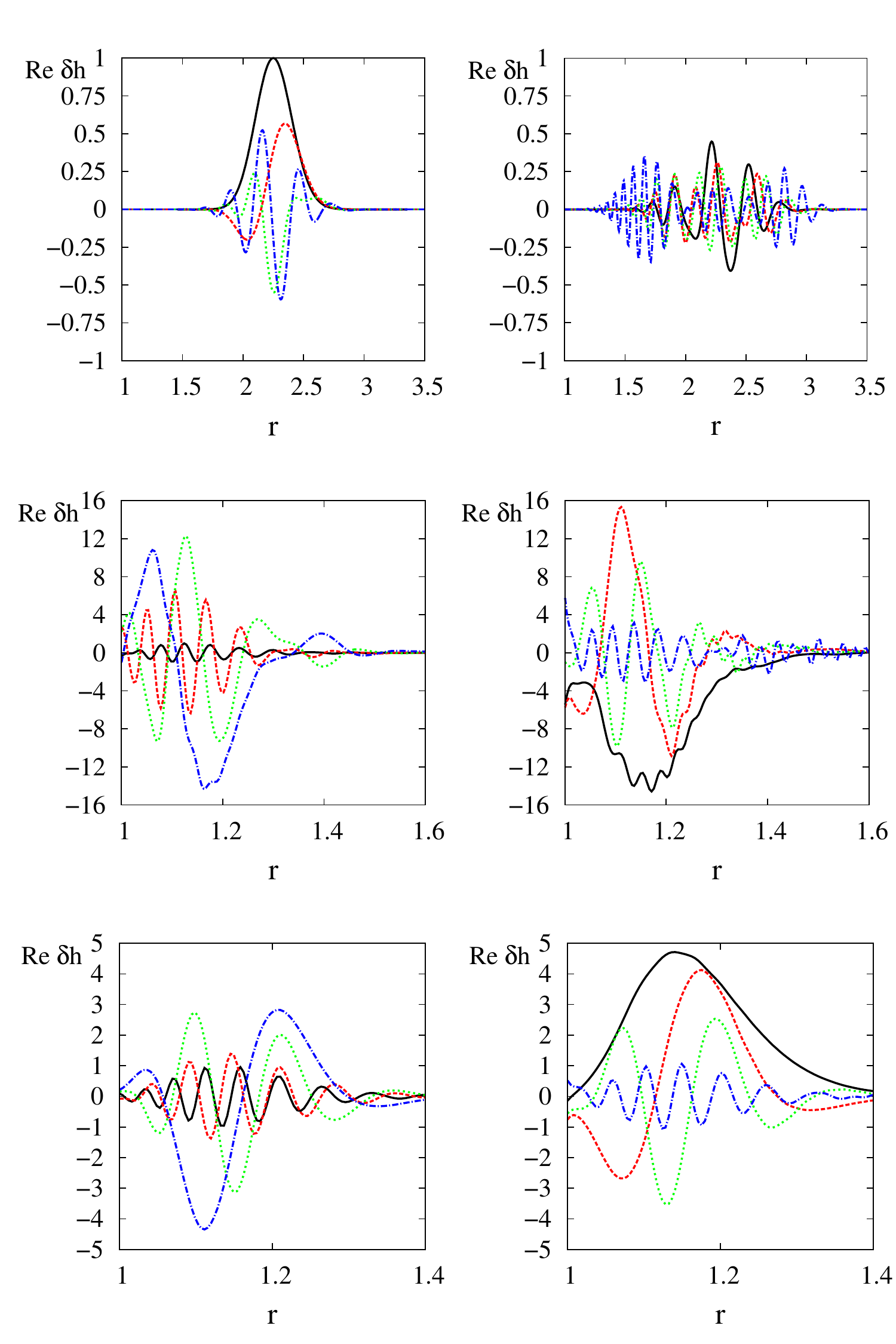}
\end{center}
\caption{In the middle and the bottom panels there are the shapes of ${\rm Re}[\delta h]$ 
for the optimised perturbation measured according to eqs. (\ref{ac_en}) and (\ref{ac_en2}), respectively. 
Background is taken as Keplerian disc with structure specified by 
eqs. (\ref{sig}, \ref{a_eq}). The optimisation time is $\tau=20$. The other parameters are $\delta=0.05$, $m=5$, $n=3/2$. 
Top panel: the shapes of ${\rm Re}[\delta h]$
for perturbation triggered as the 'single' Gaussian function (eq. \ref{gauss1}) 
and evolving in the same background.
Left-hand panels: solid, dashed, dotted and dot-dashed curves are taken 
at $t=0,5,15,20$, respectively.  Right-hand panels: solid, dashed, dotted and dot-dashed curves are taken 
at $t=25,30,35,50$, respectively. ${\rm Re}[\delta h]$ is scaled by its maximum absolute value at $t=0$ in all panels.
Find the corresponding movie at  http://xray.sai.msu.ru/$\sim$dima/movies/Evolution\_of\_perturbations.avi.zip
 } \label{fig9}
\end{figure}

Finally, we take the Keplerian disc model with the structure specified by 
eqs. (\ref{sig}, \ref{a_eq}) and determine the optimal large scale vortex employing both norms (eqs. (\ref{ac_en})
and eq. (\ref{ac_en2})). We set all parameters to the same values as in fig. (\ref{fig8}). 
The optimal growth is found to become approximately two times smaller than for the homogeneous disc model. 
The instant profiles of ${\rm Re}[\delta h]$ for optimal solutions are plotted in fig. (\ref{fig9}).
Also, we show the slices of ${\rm Re}[\delta h]$ for perturbation triggered at $t=0$ by the profile (\ref{gauss1}). 
An interested reader is invited to watch the corresponding movie, finding the reference in caption to 
the fig. (\ref{fig9}).
Comparing panels in fig. (\ref{fig9}) one can see that the incidental perturbation demonstrates a
typical wave-like evolution splitting into two waves running in the opposite directions(top panels). 
Though, it can be checked that this perturbation has non-zero potential vorticity, i.e. being a 
mixture of vortex and density wave, there is no signature of transient growth according neither of 
norms are used in this work. 
At the same time, the optimised shape of ${\rm Re}[\delta h]$ first exhibits 
the correlated transient enhancement in the vicinity of the disc inner boundary. 
According to the results of our study in the local framework, the initial optimal perturbation is
nearly identical to a pure vortex. Moreover, at $t=0$ it is indeed a leading spiral which is shrinking
due to the shear during the phase of transient growth.  
Additionally, since we look the case of large scale perturbations with 
an azimuthal wavelength much larger than the disc thickness, the excitation of density wave at the instant
of swing of the spiral is quite insignificant.  
Note that in case of scaling by the total acoustic energy (middle panels) 
the profiles of ${\rm Re}[\delta h]$ bear small scale variations at the instant of swing, see the solid curve in
the middle right-hand panel of fig. (\ref{fig9}). This is
a distinctive feature of modal solutions obtained in WKBJ approximation in thin disc, see
e.g. \citet{Kato1987}. Clearly, the radial size of these variations is dictated by the disc thickness. 
We presume that they emerge due to the contribution of oscillatory motions with frequencies
of order of the Keplerian frequency which is a hint of epicyclic deviations included in optimisation 
with norm given by eq. (\ref{ac_en}). 
Indeed, the change to norm (\ref{ac_en2}) eliminates these small scale variations producing the smooth instant profile
of the optimal perturbation at the instant of swing, see the solid curve in
the bottom right-hand panel of fig. (\ref{fig9}).
Both of optimal instant shapes strongly differ from modal solutions as well as from 
the evolution of randomly taken perturbation (top panels).

\section{Conclusions}

In this work we study the transient dynamics of linear perturbations in thin Keplerian discs. 
It is shown, that substantial non-modal growth may exist at all spatial scales including 
those when the azimuthal wavelength of perturbations is much larger than the disc thickness. 
Moreover, this remains true if the azimuthal wavelength becomes comparable to the radial 
scale in disc. In the latter situation we have been dealing with the global perturbations.    
The most reasonable way to illustrate the transient activity is to solve the optimisation problem,
finding in this way the configuration of the initial perturbations that determines
the largest possible non-stationary response of the disc.
Such optimals have been calculated for compressible perturbations in Keplerian thin disc without viscosity.

First this is done in the shearing sheet model assuming that perturbation azimuthal wavelength is small compared
to the disc radial scale but that it can be in any ratio with the disc thickness what is fixed by a parameter $R$, 
see section 3.2.1 and eq. (\ref{decoupled}) therein. Before presenting the numerical results for 
optimal shearing harmonics we consider analytically two opposite cases of small scale local perturbations, $R\ll 1$, 
and large scale local perturbations, $R\gg 1$, and derive analytical expressions for magnitude of transient growth of 
vortices in both situations, see eq. (\ref{SFH_G}) (or simplified eq. (\ref{G_lowR})) and eq. (\ref{G_highR}), respectively.  
It is important to note that in the latter case the magnitude of transient growth is $\propto 4\Omega^4/\kappa^4$ 
what suggests that vortices with $R\gg 1$ can exhibit much stronger amplification in a shear flow that approaches
a uniform specific angular momentum distribution, e.g. in the inner parts of relativistic accretion discs 
around the black holes. Besides, we formulate a condition of separability of vortices and density waves in compressible
shear flow which, in particular, leads to a condition of the validity of eq. (\ref{G_highR}).
The latter requires that the leading spiral corresponding to a vortex must be initially tightly wound, see eq. (\ref{coupled_cond}).
This restriction is particularly strong for large scale vortices, $R\gg 1$, but luckily the comparison 
of analytical and numerical growth factors of vortices indicates that actually a reasonable agreement holds up to 
$R\sim |\beta|/2$, see the comments to fig. (\ref{fig8}) in section 4.2 later on. 
We also assess the influence of non-zero viscosity on the transient growth of these large scale vortices and find that
its absolute maximum is given approximately by eq. (\ref{G_max}) what leads to estimate that in a weakly 
viscous discs with $\alpha<\delta$ the amplification of vortices exists for all possible azimuthal wavelengths 
up to the largest one, $\sim H/\delta$, see the end of section 3.2.2.

While proceeding to the optimisation in the shearing sheet model we find that generally the optimal shearing 
harmonic (SFH) is a mixture of vortex and density wave, see fig. (\ref{fig3}). 
However, the contribution of density wave sharply decreases with transition to negative $\beta$, i.e. to initially 
leading spirals.
Since for reasonable optimisation timespans, $\tau>1$, the largest optimal growth, $G(\beta)$, 
is produced by SFH with $\beta < -1$. We find that performing the optimisation for a particular $R$ 
one always gets the initial shape of the optimal SFH nearly identical to vortex having
the same potential vorticity. Apart, in the case $\beta>0$ we also find $G(\beta)>1$ what is explained by
the non-modal growth of (zero potential vorticity) density waves, see also the 4th pair of curves in fig. (\ref{fig3}).
Further, $G(R)$ by itself attains maximum value at $R\approx 1$ what is a consequence
of the density wave excitation by vortices. The excitation of density waves becomes most prominent 
for $R\approx 1$, look the analytical investigation by HP. The maximum of $G(R)$ is provided by vortex 
that swings from leading to trailing spiral approximately 1.5 times earlier than the optimisation time. 
This is explained by the fact that amplitude and growth rate of the excited density wave 
becomes sufficiently high to exceed the magnitude of another vortex which swings at the optimisation time.
The enhancement of optimal growth due to this effect is quite substantial and increases with time, look at fig. (\ref{fig5}). 
Additionally, we demonstrate that optimal SFH is indistinguishable from the analytical solution 
obtained by HP taken with the same potential vorticity, see fig. (\ref{fig4}).

Determining numerically the optimal large scale SFH, $R\gg 1$, we find an approximate agreement with 
our analytical estimates for large scale vortices, see eqs. (\ref{G_highR}) and (\ref{G_highRt}), 
fig. (\ref{fig6}) (top panel) and fig. (\ref{fig8}), as well as
discussion to them. As it has been already mentioned, this agreement holds despite 
the swing interval is comparable to the duration of growing phase for the particular values of the parameters 
used in calculations. 
Locally, for $R\gg 1$ the optimal growth (approximately, $G\propto R^{-2}$) 
is highly suppressed in comparison with its incompressible ($R\ll 1$) magnitude, see the top panel of fig. (\ref{fig6}). 
However, the situation changes when we extend our study to a global spatial scale 
taking into account the background vorticity gradient and the disc cylindrical geometry. 
We find that while $m \to 1$ the optimal growth falls down much more sharply in case of $R \ll 1$ 
rather in case of $R \gg 1$, see the bottom panel of fig. (\ref{fig6}) and fig. (\ref{fig7}). 
For example, in the particular case $m=1$, $\tau=20$ and $R=12$ the optimal growth is just 1.5 times smaller than 
its counterpart at $R\ll 1$ (i.e. formally in our disc with $\delta\gg 1$). 
Actually, the last limit corresponds to global incompressible dynamics considered by \citet{IK2001}, 
see the inviscid dependence in their fig. (1). Thus, contrary to what we have in local problem, the global 
vortices with the lowest azimuthal wavenumbers, $m=1,2...$, at least in the range $R\lesssim 10$, exhibit 
the transient growth comparable to what has been found previously in the simplified model of 
incompressible Keplerian flow. Thus, any kind of persistent source of the potential vorticity on the scales 
above the disc thickness may lead to formation and growth of global vortices providing an 
enhanced angular momentum transfer to disc periphery. At the same time, any kind of 
weak pre-existing turbulence may become such a natural sower of the potential global vorticity perturbations in disc.  
With regards to the incompressible Keplerian flow this was shown by \citet{IK2001} who introduced
the action of pre-existing turbulence as an external stochastic forcing in hydrodynamical equations. 
Importantly, they found that the coherent structures that emerge in the steady state of the perturbed disc 
are similar to the structures of global optimal perturbations sliced at the instant of swing from leading to 
trailing spirals. More precisely, the instant of swing of these optimal perturbations corresponds to the 
timespan at which the curve of optimal growth attains its maximum provided that disc is viscous. 
Since in this work we show, that the transient growth of global vortices preserves its strength in 
thin compressible discs, it is tempting to suggest that they are responsible for an additional accretion 
stress on the scales well above the disc thickness. As we discuss at the end of section 4.1, the recent 
results of global MHD simulations of disc turbulence probably provide some evidence for that by detecting
the non-local accretion stress. 

At last, let us highlight the results of a more methodical and technical nature. 
First, we find that non-modal dynamics in the perturbed flow measured by the total acoustic energy of perturbations 
leads to a distinctive feature in the curve of optimal growth as a function of time.
Particularly, at the time intervals shorter that one rotational period at the inner disc boundary 
there is a quick raise of optimal growth up to a flat segment at magnitude which approximately 
equals to the squared ratio of epicyclic frequency for rigid rotation to epicyclic frequency for
the given shear. This feature is independent on the value of azimuthal wavenumber and we suggest that it
is related to the existence of epicyclic motions in disc. In order to support this conclusion
we supply this work with the Appendix A where the non-modal growth of axisymmetric compressible perturbations
is considered. We carry out an analytical treatment and find that the optimals are the standing density waves with
the optimal growth given by eq. (\ref{G_an}). Standing wave is the configuration that 
naturally leads to oscillations of the acoustic energy integrated over its wavelength. 
Then, since in the shear flow perturbations of centrifugal and centripetal forces acting 
on a displaced fluid particle are not balanced their non-zero difference serves as the source of
the kinetic energy for the fluid particle. Thus, one obtains an additional, inertial mechanism of 
non-modal growth in differentially rotating flows. It dominates in the limit of small sound speed 
and long wavelength and yields an optimal growth given by eq. (\ref{G_an2}). 
Note that this is not the case for non-axisymmetric perturbations of either type since 
their dynamics is mostly dictated by the perturbed pressure gradients and is subject to the classical lift-up effect. 
However, it is important to note that these findings about the optimal axisymmetric perturbations 
are true as long as the acoustic energy is employed to measure the compressible dynamics. 
It turns out that we are able to choose norm of perturbations in such a way that the optimal growth 
identically equals to unity in axisymmetric case, see eq. (\ref{ac_en2}).
This is an energy-like quantity which becomes equivalent to the canonical energy in case of axisymmetric perturbations. 
It allows to exclude the degenerate limit of the non-modal growth associated with general oscillatory solution for 
epicyclic motions in disc and stay solely with a transient growth. 
Consequently, the optimal growth $G\to 1$ as $R\to\infty$ (or $m=0$) and the slices of the particular optimal solutions 
acquire a smooth shape, see fig. (\ref{fig9}) for comparison of optimal perturbations measured by two alternative norms.

Second, we describe a useful optimisation technique that has not been applied in the astrophysical discs
theory before. It allows to study the transient effects directly by solving
the Cauchy problem for perturbations in the framework of an iterative scheme when 
the basic set of equations is advanced forward in time whereas the adjoint set of equations
is advanced backward in time. This is done
without referring to modal solutions what can
be quite an involved task, especially in complex flows. For example,
\citet{ZhSh2009} and \citet{RZh2012} treated the optimals in the form of finite linear combinations
of neutral acoustic modes in a quasi-Keplerian torus.
The obtained optimal perturbation corresponds to a wave packet
localised initially in the vicinity of the outer boundary of torus and moving towards the inner boundary.
At the moment of the reflection from the inner boundary the total acoustic energy attains its
maximum. The constituent modes are phased in such a way that the shape of wave packet corresponds to the highest
possible energy profit at this maximum for the given time interval.
However, these results are not robust in the sense that to make a decisive conclusions about the transient
dynamics one has to cover all possible combinations of modes including those with corotational
and Lindblad resonances inside the flow. Note that neutral and damping modes with corotation inside the
flow must also be considered what is an involved task since one has to extend calculations
in complex plane according to the Lin's rule. This is a complex problem especially if one would like to
investigate the influence of stratification and even baraclinity of the flow on its capability for
transient behaviour.
On the contrary, in this work we have avoided all such issues since we have not been obliged to obtain 
modal solutions in order to study the optimal growth in disc. 

Using the arguments from the operator theory we show that the variational technique may be applied 
to the non-stationary accretion flows when one has to solve differential equations 
with time-dependent coefficients. In this case the spectral problem can not the formulated at all. 
However, the transient growth and the optimal perturbations may exist in such a flow and could be 
determined using corresponding iterative procedure. 
Also note that the variational technique may be applied to the
non-linear problems as well. At last, let us point out that it can be used to find the whole set of singular
vectors. In order to do that, one can apply the same procedure
in the functional subspace normal to the one spanned by previously obtained singular vectors.

\section*{Acknowledgements}
We are grateful to P.B. Ivanov for useful conversations and critical comments.
We thank anonymous referee for substantial remarks that helped to improve the manuscript.
The study was supported
in part by grant RFBR 12-02-00186, in part by grant RFBR-NSFC 14-02-91172, 
in part by M.V.Lomonosov Moscow State University Program of Development and
in part by programme 22 of the presidium of RAS.

\bibliographystyle{mn2e}
\bibliography{text}

\appendix

\section{Particular case of axisymmetric perturbations measured by their acoustic energy}

In comments to fig. (\ref{fig7}) we discuss that for axisymmetric 
perturbations, i.e. with $m=0$, measured by their total acoustic energy $G=4$ provided that the shear is Keplerian. 
The question should be addressed what is the reason 
for non-modal growth in that situation when the classical lift-up effect does not work. 
In order to discuss this issue in detail we would like to carry out an analytical investigation. 
It is not difficult to do in the shearing sheet approximation. 
The general solution to equations (\ref{sonic_sys1_sh}-\ref{adj_sonic_sys3_sh}) with $k_y=0$
reads as follows
\begin{equation}
\label{sol_axisymm}
\begin{array}{l}
\hat u_x = C_1 e^{{\rm i}\sigma t} + C_2 e^{-{\rm i}\sigma t}\\
\hat u_y = \frac{{\rm i} \kappa^2}{2\Omega_0^2 \sigma} 
\left (  C_1 e^{{\rm i}\sigma t} - C_2 e^{-{\rm i}\sigma t} \right )  + \frac{k_x}{2} C_3 \\
\hat W = -\frac{k_x}{\sigma} 
\left ( C_1 e^{{\rm i}\sigma t} - C_2 e^{-{\rm i}\sigma t} \right ) - {\rm i} \, C_3
\end{array}
\end{equation}

\begin{equation}
\label{adj_sol_axisymm}
\begin{array}{l}
\hat {\tilde u}_x = K_1 e^{{\rm i}\sigma t} + K_2 e^{-{\rm i}\sigma t}\\
\hat {\tilde u}_y = \frac{2 {\rm i} }{\sigma} 
\left (  K_1 e^{{\rm i}\sigma t} - K_2 e^{-{\rm i}\sigma t} \right )  + \frac{k_x}{2} K_3 \\
\hat {\tilde W} = -\frac{k_x}{\sigma} 
\left ( K_1 e^{{\rm i}\sigma t} - K_2 e^{-{\rm i}\sigma t} \right ) - \frac{{\rm i}\,\kappa^2}{4\Omega_0^2} K_3
\end{array}
\end{equation}
where $(C_1,C_2,C_3)$ and $(K_1,K_2,K_3)$ are complex constants 
to be determined by the iterative procedure and the dimensionless $\sigma^2 = \kappa^2/\Omega_0^2 + k_x^2$.

The analytical expression for norm of the perturbations state vector, $||{\bf q}||$, 
is derived using eq. (\ref{loc_norm}). It has the form
\begin{eqnarray} 
||{\bf q}||^2 = (|C_1|^2 + |C_2|^2) \left ( 1 + \frac{\kappa^2 s^{-1}/\Omega_0^2 + k_x^2}{\sigma^2} \right ) + \nonumber \\
|C_3|^2 \left ( 1 + \frac{k_x^2}{4} \right ) + \nonumber \\
2 \,{\rm Re} \left [ C_1 C_2^* e^{2{\rm i}\,\sigma t} \right ] 
\left ( 1 - \frac{\kappa^2 s^{-1}/\Omega_0^2 + k_x^2}{\sigma^2} \right ) + \nonumber \\
2 \frac{k_x}{\sigma}\, (s^{-1}-1)\, {\rm Re} \left [  {\rm i}\, C_3^* \left ( C_1 e^{{\rm i}\,\sigma t} - 
C_2 e^{-{\rm i}\,\sigma t} \right ) \right ], \label{loc_norm_an} 
\end{eqnarray}
where the asterisk denotes complex conjugation and $s \equiv 4\Omega_0^2 / \kappa^2$ is a parameter
that characterises the shear magnitude, e.g. for rigid rotation $s=1$ and for Keplerian rotation 
$s=4$. 
Looking at eq. (\ref{loc_norm_an}) one concludes that whenever $s \neq 1$ $||{\bf q}||$ becomes a time-dependent
quantity. 
Now, to obtain optimal perturbations corresponding to some moment $t=\tau$ we can proceed in two 
different ways. 

\begin{itemize}

\item

Let us construct an iterative scheme for the coefficients $C_1, C_2, C_3$. Specifically, matching 
the state and the adjoint vectors at $t=\tau$, ${\bf \tilde q}_p(\tau)={\bf q}_p(\tau)$, with the help of 
eqs. (\ref{sol_axisymm}, \ref{adj_sol_axisymm}) we express $(K_1,K_2, K_3)_p$ through $(C_1, C_2, C_3)_p$, where
$p$ is a number of iteration. Then, matching ${\bf q}_{p+1}(0)={\bf \tilde q}_p(0)$ and dividing 
${\bf q}_{p+1}(0)$ by its own norm according to eq. (\ref{loc_norm_an}) we finally 
get the recursive relations for coefficients $(C_1, C_2, C_3)_{p+1}$ expressed in terms of 
$(C_1, C_2, C_3)_{p}$ known from the previous step. Starting from an arbitrary set of $(C_1, C_2, C_3)$
we get a converging sequence of coefficients which gives us an optimal perturbation corresponding
to $t=\tau$ and the value of the optimal growth, $G(\tau)$.
 
\item

Let us determine the maximum of $g(t) \equiv ||{\bf q}(t)||^2 / ||{\bf q}(0)||^2$ considering it as function of coefficients 
$C_1, C_2, C_3$. This will give us an optimal growth curve. Below we describe the details of the analytical 
derivation of $G(\tau)$ in case of small shear ($s\approx 1$) and discuss the physical reasons for non-modal 
growth of axisymmetric perturbations.
\end{itemize}

\subsection{Optimal growth of axisymmetric perturbations in a rotating flow with a small shear}

\begin{figure*}
\begin{minipage}[h]{0.48\linewidth}
\includegraphics[width=0.88\linewidth]{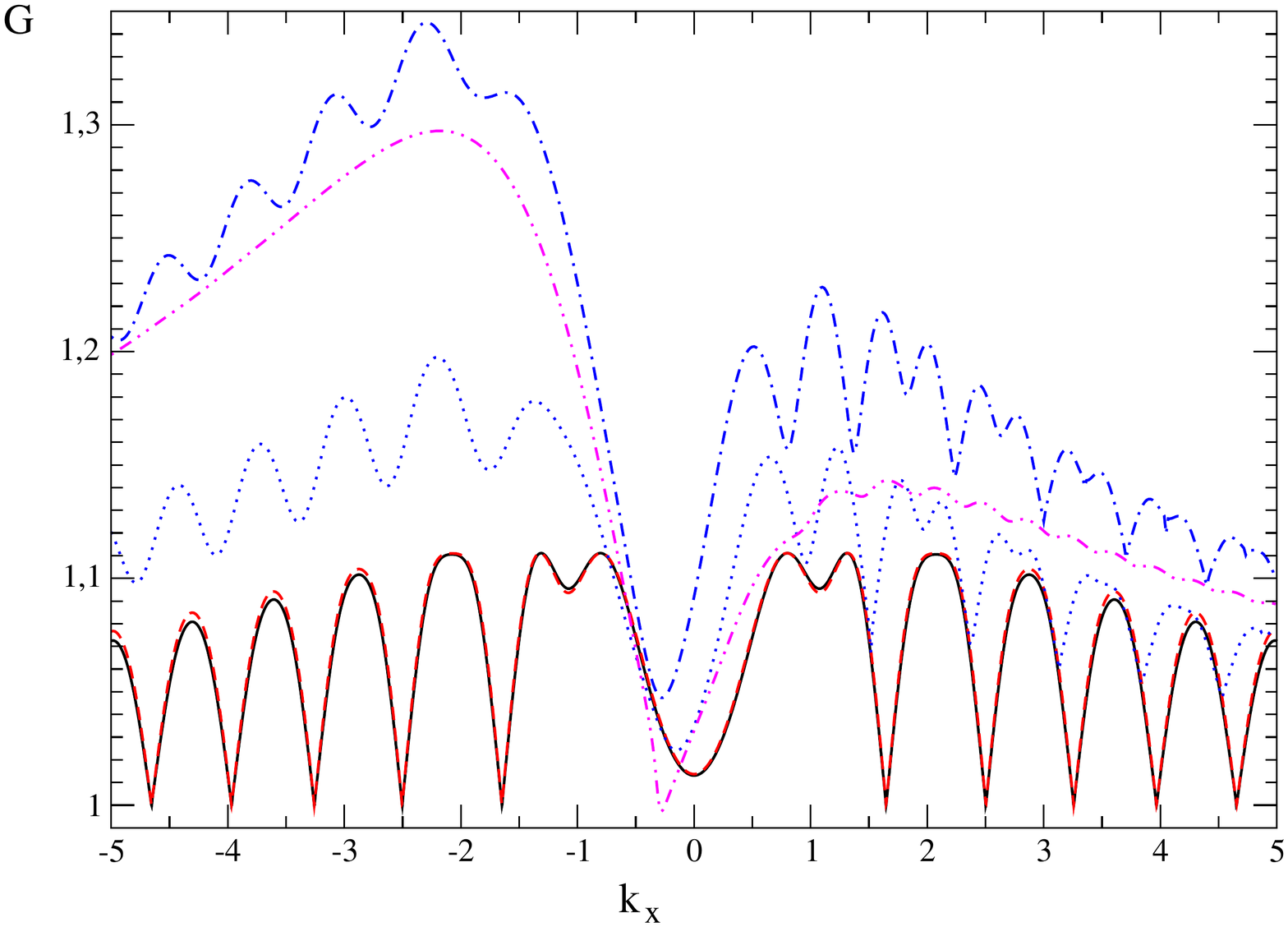} 
\end{minipage}
\begin{minipage}[h]{0.48\linewidth}
\includegraphics[width=0.88\linewidth]{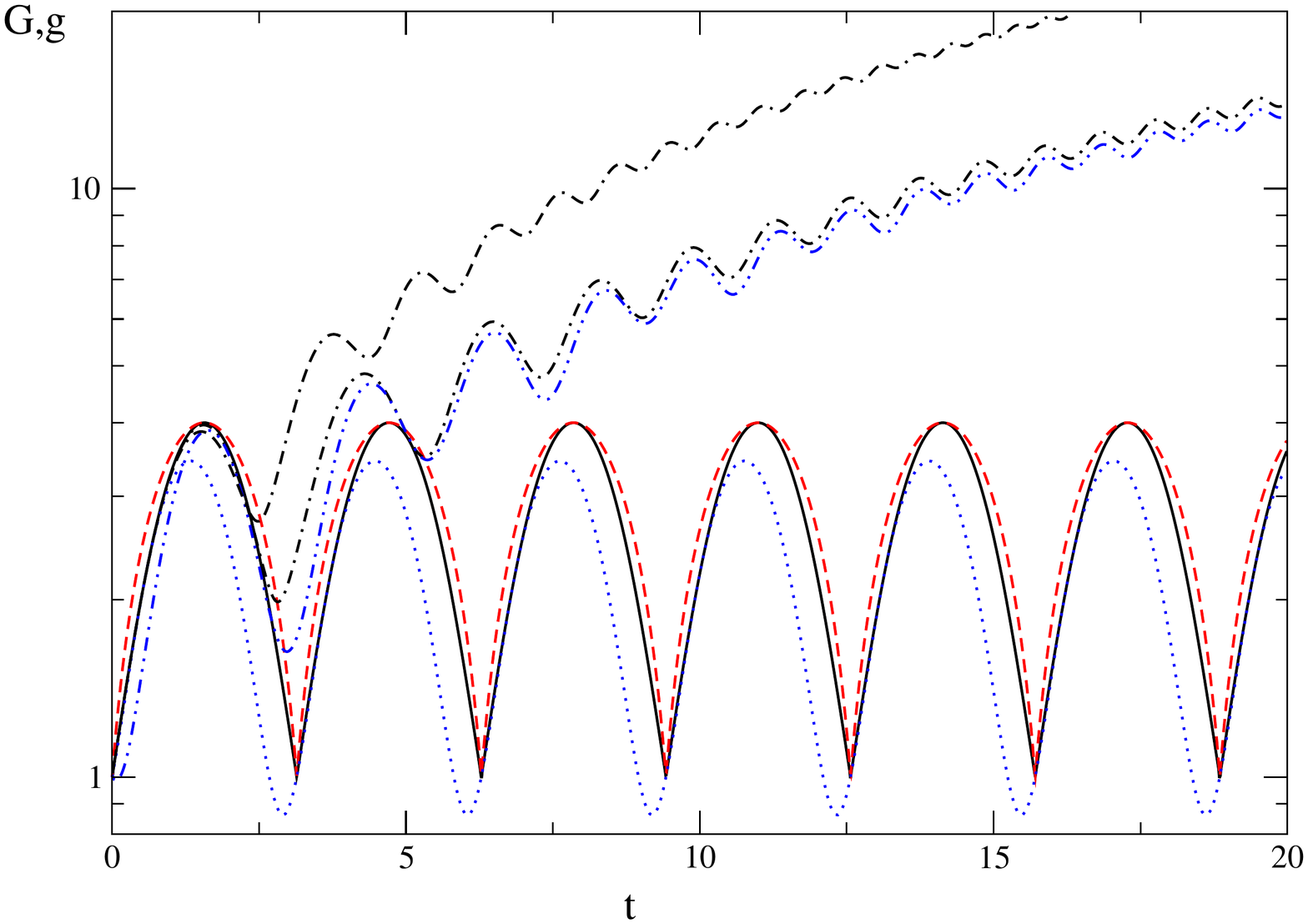} 
\end{minipage}
\caption{Optimal growth in the shearing sheet approximation. Left-hand panel: $G$ as function of $k_x$ for 
$\tau=10$ and $q=0.2$.
Solid curve is obtained for $k_y=0$ iterating the analytical solutions (\ref{sol_axisymm},\ref{adj_sol_axisymm}), 
dashed curve represents the analytical expression (\ref{G_an}),
dotted and dot-dashed curves are obtained applying the iterative scheme with equations 
(\ref{sonic_sys1_sh}-\ref{adj_sonic_sys3_sh}) and setting $k_y=0.125,0.25$, respectively, whereas 
dot-dot-dashed curve is obtained applying the iterative scheme with equations 
(\ref{sonic_sys1_sh}-\ref{sonic_sys3_sh}, \ref{adj_sonic_sys1_sh_2}-\ref{adj_sonic_sys3_sh_2}) 
and setting $k_y=0.25$.
Right-hand panel: $G$ as function of $t$ for $k_x=0$ and $q=1.5$.
Solid curve is obtained for $k_y=0$ iterating the analytical solutions (\ref{sol_axisymm},\ref{adj_sol_axisymm}), 
dashed curve represents the analytical expression (\ref{G_an2}),
dotted curve gives $g$ for optimal perturbation with $\tau=10$.
Dot-dashed and dot-double-dashed curves are obtained applying the iterative scheme with equations 
(\ref{sonic_sys1_sh}-\ref{adj_sonic_sys3_sh}) and setting $k_y=0.125,0.25$, respectively.
Dot-dot-dashed curve gives $g$ for $k_y=0.125$ and $\tau=5$.
} \label{fig10}
\end{figure*}

Let us suppose that $s = 1+\epsilon > 1$, where $\epsilon<<1$. 
Also we notice that if one regards $(C_1,C_2,C_3)$ as vectors with components $C_i=(X_i,Y_i)$
in some Cartesian reference frame, then 
$
C_{1} C_{2,3}^* = |C_{1}| |C_{2,3}| (\cos \psi_{1,3} - {\rm i}\, \sin \psi_{1,3})
$
and
$
C_{2} C_{3}^* = |C_{2}| |C_{3}| (\cos \psi_{2} - {\rm i}\, \sin \psi_{2}),
$
where $\psi_{1,2,3}$ are the angles between vectors $(C_1,C_2)$, $(C_2,C_3)$ and $(C_1,C_3)$, respectively. 
Note that by definition $\psi_3 = \psi_1 + \psi_2$.

The equalities above allow us to write the growth factor of some perturbation, $g(t,k_x)$, expanding eq. (\ref{loc_norm_an})
over the small $\epsilon$ and retaining the linear term only
\begin{eqnarray}
\label{g_an_1}
( g - 1 ) \epsilon^{-1} = \frac{2}{2(A_1^2+1) + A_2^2 (1+k^2)} \times\nonumber \\
\left \{ 
\frac{A_1}{1+k^2} (\cos(2\sigma t-\psi_1) - \cos \psi_1) + \right . \nonumber \\
\frac{|k|A_2}{(1+k^2)^{1/2}} \, [\, A_1 ( \sin(\sigma t - \psi_1 - \psi_2) +   
\sin (\psi_1+\psi_2)) + \nonumber \\
\sin (\sigma t + \psi_2) - \sin \psi_2  \,]\,
\Big\},
\end{eqnarray}
where $k \equiv k_x/2$ and $A_{1} \equiv |C_{1}|/|C_2|$, $A_{2} \equiv |C_{3}|/ |C_2|$.

We are going to determine the maximum of $g$ provided that $t$ and $k_x$ are fixed. 
At first, let us set $A_1=1$.
After that it is straightforward to obtain the values of $\psi_{1,2}$ and $A_2$ corresponding to 
the maximum of eq. (\ref{g_an_1}) as a function of $\psi_1,\psi_2$ and $A_2$. We have
\begin{eqnarray}
\label{g_an_2}
\psi_1 = -2\psi_2, \quad
\quad \sin(\psi_2+\sigma t/2) = \nonumber\\ \frac{  
A_2 |k|(1+k^2)^{1/2} - (A_2^2 k^2(1+k^2)+32\cos^2 (\sigma t/2)  )^{1/2} } { 8 \cos(\sigma t/2) }, \nonumber \\
A_2=\frac{2|k|}{(1+k^2)^{1/2} (k^2 + 2\cos^2 (\sigma t/2))^{1/2}}
\end{eqnarray}

Finally, it is not difficult to check that eqs. (\ref{g_an_2}) turn into identities 
the conditions of maximum of $g$ (from eq. (\ref{g_an_1}) ) over the varying $A_1$. 
Thus, we make sure that $A_1=1$ along with eqs. (\ref{g_an_2}) 
correspond to the optimal growth case indeed.

Substituting eqs. (\ref{g_an_2}) along with $A_1=1$ into eq. (\ref{g_an_1}) yields
\begin{equation}
\label{G_an}
(G-1)\, \epsilon^{-1} = 2 \, |\sin(\sigma t/2)| \, \frac{(k^2 + \cos^2 (\sigma t/2) )^{1/2}}{1+k^2}
\end{equation}

Finally, we notice that setting $k=0$ (what corresponds to $k_x\to 0$)
eq. (\ref{G_an}) turns into a simple expression for $G$, explicitly
\begin{equation}
\label{G_an2}
(G-1)\, \epsilon^{-1} = |\sin(\sigma t)|
\end{equation}

Thus, $G$ oscillates both with time and radial wavenumber. 

\subsection{What causes non-modal growth of axisymmetric perturbations}

We plot the results of calculations in fig. (\ref{fig10}).
For strictly axisymmetric perturbations we show curves of optimal growth obtained by iterative scheme for
coefficients $C_1,C_2,C_3$ and based 
on the analytical solutions (\ref{sol_axisymm},\ref{adj_sol_axisymm}). As can be seen, the analytical 
expression (\ref{G_an}) gives the profile of $G(k_x)$ which is in a good agreement with the iterative one
for the case of small shear (left-hand panel in fig. \ref{fig6}). Also, the analytical expression (\ref{G_an2})
describes well the behaviour of $G(\tau)$ for the case of long wavelength perturbations, $k_x\to 0$
(the right-hand panel in fig.\ref{fig6}). 
Note that in the latter case the shear is not small and is set to its Keplerian value, i.e. $s=4$. 
Clearly, the optimal growth always attains the value of $s$. It is symmetric 
with respect to change $k_x\to -k_x$ and gradually tends to $1$ while $|k|\to \infty$ what 
corresponds to an ordinary sound waves in the absence of shear. To illustrate the transition
to non-axisymmetric optimal perturbations we add curves corresponding to azimuthal wavenumbers
$k_y=0.125,0.25$\footnote{We also plot $G(k_x)$ with $k_y=0.25$ for perturbations
measured by eq. (\ref{loc_norm_2}) that do not exhibit any kind of non-modal growth in case $k_y=0$.}, 
i.e. to $R=12$ and $R=6$, respectively. 
Non-modal growth is no more limited by the value of $s$ and becomes larger 
as $R$ decreases. The shape of $G(\tau)$ transforms to its familiar form 
(cf. fig. (\ref{fig2}), the dot-dashed-dashed curve).
Specifically, $G(k_x)$ becomes asymmetric with respect 
to change $k_x\to -k_x$ getting larger in the domain of negative $k_x$, since 
the non-modal growth appears in this case due to the leading spirals being shrunk by the shear. 

The variant of iterative scheme used for axisymmetric perturbations as well as 
the results of the analytical consideration made in the previous section show that 
optimal perturbations are the standing axisymmetric inertial-acoustic waves. Indeed, we have got
that in the optimised solution $|C_1|=|C_2|$, i.e. the optimal perturbation is a combination of 
two monochromatic waves running in opposite directions. 

The limit of $k_x\to 0$ gives an especially simple version of the optimal perturbations.
These are nothing but cophased epicyclic oscillations in the rotating flow.  
The velocity perturbations obey the following equations
\begin{equation}
\label{epic_1}
\frac{\partial u_x}{\partial t} = 2\Omega_0 u_y,
\end{equation}
\begin{equation}
\label{epic_2}
\frac{\partial u_y}{\partial t} = -(2-q)\Omega_0 u_x.
\end{equation}
Let us change to the Lagrangian approach and rewrite eqs. (\ref{epic_1},\ref{epic_2}) 
in terms of the Lagrangian displacement, $\mbox{\boldmath$\xi$}$, associated with the particular fluid particle. 
In this simple case the Lagrangian velocity perturbations are $w_x=u_x$ and $w_y=u_y-q\Omega_0 \xi_x$ 
and $\mbox{\boldmath$\dot\xi$} = {\bf w}$, see e.g. \citet{LOstr67}. The Lagrangian time derivative is denoted by dot.
We get 
\begin{equation}
\label{xi_1}
\ddot \xi_x = 2\Omega_0 (\dot \xi_y + q\Omega_0  \xi_x),
\end{equation}
\begin{equation}
\label{xi_2}
\ddot \xi_y = -2\Omega_0 \dot\xi_x.
\end{equation}
Looking at eqs. (\ref{xi_1},\ref{xi_2}) we see that along with the Coriolis force acting on the particle 
there is an additional conservative force, ${\bf f}$, with a non-zero radial component, $f_x=2q\Omega_0^2 \xi_x$, which is 
proportional to the shear. The presence of ${\bf f}$ is explained as follows.
In absence of perturbations of pressure gradient the dynamics of fluid particle is determined by the difference 
between a perturbation of the centrifugal force, $\Omega_0^2 \xi_x$, and a perturbation of the centripetal force, 
$2q\Omega_0^2 \xi_x+\Omega_0^2\xi_x$, which vanishes only in case of rigid rotation.

Thus, there is an ``energy'' integral of motion 
\begin{equation}
\label{xi_int}
E = w^2/2 - q\Omega_0^2 \xi_x^2={\rm const}. 
\end{equation}
Eq. (\ref{xi_int}) indicates that the kinetic energy of the fluid particle, $E_k=w^2/2$, changes due to the work 
done by ${\bf f}$. This work is positive when the particle moves away from its unperturbed position. 
So ${\bf f}$ is a destabilising force. In a limiting case $q=2$ the frequency of radial 
oscillations (i.e. the epicyclic frequency $\kappa$) vanishes and the motion becomes marginally stable. 
At the same time, $E_k$ can increase infinitely together with $\xi_x$.
Thus, $E_k$ remains constant only in the absence of shear since the Coriolis force does not do work.

A remark on the ``energy'' integral (\ref{xi_int}) was made e.g. by \citet{GT80} (cf. their eq. (35)), 
where they represented the epicyclic motion problem from the mechanical point of view. 
The epicyclic trajectory of the fluid particle is an ellipse elongated along the radial direction 
so when $\xi_x=0$ the azimuthal component of the velocity, $\dot\xi_y$, vanishes. 
This allows one to find that $\xi_x = -\dot\xi_y / (2\Omega_0)$ 
integrating eq. (\ref{xi_2}). Using eq. (\ref{xi_int}) we find that 
the ratio of maximum and minimum values of the kinetic energy during one orbit equals to 
$s=4\Omega_0^2/\kappa^2$. This fact explains the non-modal growth of axisymmetric perturbations
with $k\to 0$, see the right-hand panel in fig. (\ref{fig10}). 
When $k\neq 0$ the standing epicyclic waves are modified by sonic component, see the left-hand panel in fig. (\ref{fig10}).
At last, when $k_y\neq 0$ they become shrunk by the shear what causes an additional 
enhancement of the kinetic energy of perturbations due to the lift-up mechanism, see both panels in 
fig. (\ref{fig10}).

\section{Incompressible perturbations}

\subsection{Optimisation on a global spatial scale}

In case of planar motions it is convenient to rewrite the dynamical equations for perturbations
of vorticity, $\delta\mbox{\boldmath$\omega$} = \nabla\times \delta{\bf v}$, and 
stream function, $\delta\mbox{\boldmath$\psi$}$, defined through 
$\delta \bf{ v} = \nabla\times \delta\mbox{\boldmath$\psi$}$. 
The stream function fully describes the velocity field because the latter is solenoidal
in case of incompressible dynamics. Since both $\delta\mbox{\boldmath$\omega$}$ and $\delta\mbox{\boldmath$\psi$}$ 
have the only one non-zero component along the $z$ axis we omit the subscript 'z' below.

Taking the curl of eq. (\ref{orig_sys1}) we get for the azimuthal Fourier harmonic of perturbations
\begin{equation}
\label{vort_sys1}
\frac{\partial \delta\omega}{\partial t} = -im\Omega \delta\omega - 
\frac{im}{r}\frac{d}{dr}\left ( \frac{\kappa^2}{2\Omega} \right ) \, \delta\psi,
\end{equation}
\begin{equation}
\label{vort_sys2}
\delta\omega = \frac{m^2}{r^2} \psi - \frac{1}{r} \frac{\partial}{\partial r} \left ( r\frac{\partial \delta\psi}{\partial r}\right )
\end{equation}

Clearly, eq. (\ref{vort_sys1}) is derived by constructing a combination 
$-\frac{im}{r} \times (\ref{sys})_1 + 
\frac{1}{r}\frac{\partial}{\partial r} \left ( r \times (\ref{sys} )_2 \right )$,
where by $(\ref{sys})_1$ and $(\ref{sys} )_2$ we denote the first and the second equation in the set (\ref{sys})
taken with ${\bf A}$ in the form of eq. (\ref{A_expl}).
Note that substituting the modal partial solutions $\propto exp(-i\sigma t)$ into the set 
(\ref{vort_sys1}, \ref{vort_sys2}) one gets the well-known Rayleigh equation for $\delta \psi$ 
that poses the two-dimensional spectral problem in the rotating shear flow provided 
appropriate boundary conditions are imposed. The Rayleigh equation yields an inflexion point 
criterion for spectral stability, see e.g. \citet{LL}.

The norm of the state vector is given by its total kinetic energy,
\begin{equation}
\label{vort_en}
||{\bf q}||^2 = \pi \int \left ( |\delta v_r|^2 + |\delta v_\varphi|^2 \right ) r\, dr,
\end{equation}
and here we assume the uniform surface density. 

To derive the adjoint equation for adjoint vorticity, $\tilde \omega$, 
we construct exactly the same combination of the first and of the second 
equation in the system (\ref{adj_sys}) taken with ${\bf A^\dag}$ in the form of eq.~(\ref{A_adj_expl}) now.

We get the following result
\begin{equation}
\label{vort_sys3}
\frac{\partial\delta\tilde\omega}{\partial t} = -im\Omega \delta\tilde\omega + 
2im r \frac{d\Omega}{dr} \frac{\partial}{\partial r} \left ( \frac{\delta\tilde\psi}{r} \right )
\end{equation}

Again, eq. (\ref{vort_sys3}) must be solved together with the relation (\ref{vort_sys2}) where 
the adjoint quantities, $\tilde\omega$ and $\tilde\psi$, must be substituted.

After all, the state vector ${\bf q}=\{\delta\omega\}$ contains solely the Eulerian perturbation of vorticity
and operator $\bf A$ in eq. (\ref{sys}) may be expressed in the form (cf. eq. (9) of \citet{IK2001})

\begin{equation}
\label{A_expl_vort}
-im\Omega - 
\frac{im}{r}\frac{d}{dr}\left ( \frac{\kappa^2}{2\Omega} \right ) \, (\nabla^2)^{-1}, 
\end{equation}
the adjoint operator ${\bf A}^\dag$ reads
\begin{equation}
\label{A_adj_expl_vort}
im\Omega - 
2im r \frac{d\Omega}{dr} \frac{\partial}{\partial r} \left ( \frac{(\nabla^2)^{-1}}{r} \right ),
\end{equation}
where the differential operator $(\nabla^2)^{-1}$ is the inverse of the $\nabla^2$ operator 
which specifies the direct relation between the vorticity and the stream function perturbations in eq. (\ref{vort_sys2}).
The inverse relation given by $(\nabla^2)^{-1}$ is well-defined provided that the appropriate 
boundary conditions are imposed. 
\\
\\
For incompressible perturbations when ${\bf A}$ and ${\bf A}^\dag$ are given by 
eqs. (\ref{A_expl_vort}) and (\ref{A_adj_expl_vort}), respectively,  
we choose another numerical scheme since the type of 
differential equations changes. We use two meshes shifted for $\Delta t/2$ relative to each other along
the time axis. Then, to evaluate the vorticity at each time slice we invert the set of difference 
equations connecting vorticity and stream function according to eq. (\ref{vort_sys2}).
To close this set of difference equations we require perturbation of
radial velocity to vanish at the boundaries.

\subsection{Optimisation in a shearing sheet model}

Locally, the general initial value problem for incompressible
perturbations has an exact analytical solution. This was first shown by \citet{Lom1988} who
also examined two-dimensional perturbations in the disc plane changing to the comoving 
Cartesian coordinates and considering particular SFH.

In the limit $a_* \to\infty$ (i.e. $k_y, k_x \gg 1$) the set (\ref{sonic_sys1_sh}-\ref{sonic_sys3_sh}) gives that
each SFH of the radial velocity perturbation obeys the following ODE (we omit the prime after the dimensionless time)
\begin{equation}
\label{SFH_eq}
\frac{d \hat u_x}{d t} +
2q \frac{\beta + q t}{1 + (\beta+q t )^2} \, \hat u_x = 0.
\end{equation}
In this way we get a simple solution of eq. (\ref{SFH_eq})
\begin{equation}
\label{SFH_sol}
\hat u_x(t) = \hat u_x(0) \, \frac{\beta^2 + 1}{(\beta + q t)^2 + 1}
\end{equation}
which of course can be reproduced also from eq. (\ref{vort_u_x}) in the limit $k_y\gg 1$.

An incompressible relation between $\hat u_x$ and $\hat u_y$
yields the SFH energy density evolution having exactly the same form as in eq. (\ref{SFH_sol})
\begin{equation}
\label{SFH_g}
g(t) = \frac{\hat u^2(t)}{\hat u^2(0)} =
\frac{\hat u_x^2(t)+\hat u_y^2(t)}{\hat u_x^2(0)+\hat u_y^2(0)}
 = \frac{\beta^2 + 1}{(\beta + q t)^2 + 1}.
\end{equation}

Now, we treat eq. (\ref{SFH_g}) as a function of $\beta$. For a fixed value
of $t=\tau$ it attains maximum at
$
\beta = 1/2 \,(-q \tau - ((q \tau)^2+4)^{1/2})
$
and the maximum value of eq. (\ref{SFH_g}) is the optimal growth defined as maximised $\mathcal{G}$, thereby in the present
idealised case
\begin{equation}
\label{SFH_G}
G(\tau) = \frac{ (q \tau)^2 + q \tau [(q \tau)^2+4]^{1/2} +4 }
{ (q \tau)^2 - q \tau [(q \tau)^2+4]^{1/2} +4 }
\end{equation}
For large times, $q \tau \gg 1$, eq. (\ref{SFH_G}) gives $G\approx (q \tau)^2$
what recovers eq. (\ref{G_lowR}).

Let us employ an iterative procedure for local vortex perturbations.
We have to find a solution of the adjoint set of equations (\ref{adj_sonic_sys1_sh}-\ref{adj_sonic_sys3_sh}) 
taken in the limit $a_* \to\infty$ (i.e. $k_y, k_x \gg 1$).  

It is not difficult to show that this leads to a trivial equation for SFH of $\tilde u_x$
\begin{equation}
\label{adj_SFH_eq}
\frac{d\hat {\tilde u}_x}{d t} = 0
\end{equation}

Now, omitting the steps (\ref{edge_terms1}) and (\ref{edge_terms2}) of the iterative procedure,
which are formal for the linear problem,
we get a factor
\begin{equation}
\label{opt_factor}
\left [ \frac{\beta^2 + 1}{(\beta + q \tau)^2 + 1} \right ]^p
\end{equation}
in front of an arbitrary initial profile of $\hat u_x$, $\hat u_x^{in}(\beta,t=0)$, which
is used to launch the procedure. $p$ is a natural number that equals to the number of iterations.
With a proper renormalisation applied when $p\to\infty$ factor (\ref{opt_factor}) discards
all SFHs from $\hat u_x^{in}(\beta,t=0)$ except the optimal one that
corresponds to a maximum of (\ref{opt_factor}) as a function of $\beta$.
With this obtained, we evidently mimic the optimal growth profile (\ref{SFH_G}).

It is instructive to come to the same conclusion considering the local limit of the sets 
(\ref{vort_sys1}, \ref{vort_sys2}) and (\ref{vort_sys3}, \ref{vort_sys1}) or, equivalently, 
taking the curl of eqs. (\ref{sonic_sys1}, \ref{sonic_sys2})  and 
the curl of eqs. (\ref{adj_sonic_sys1}, \ref{adj_sonic_sys2}) (implying the incompressible limit again) 
for state and adjoint quantities, respectively. 
Then, changing to the shearing coordinates we obtain for $\hat \omega$ 

\begin{equation}
\label{SFH_eq_vort}
\frac{d\hat \omega}{dt} = 0,
\end{equation}
whereas  $\hat{\tilde\omega}$ obeys the following equation

\begin{equation}
\label{adj_SFH_eq_vort}
\frac{d\hat{\tilde\omega}}{d t} -
 2q \frac{\beta + q t}{1 + (\beta+q t)^2} \, \hat{\tilde \omega} = 0,
\end{equation}

Eq. (\ref{SFH_eq_vort}) represents the law of vorticity conservation what is expected for a perfect incompressible fluid. 
However, we conclude that the adjoint vorticity does not conserve as far as there is non-zero shear in the flow. 
Comparing eq. (\ref{adj_SFH_eq_vort}) with eq. (\ref{SFH_eq}) one finds that they are identical to each other
up to a change $t \to -t$. But according to the iterative procedure eqs. (\ref{adj_SFH_eq_vort})
and (\ref{SFH_eq}) must be integrated in opposite directions in time what leads us again to a factor 
(\ref{opt_factor}) after $p$ iterations. 

Let us also note that in the limit $a_*\to\infty$ an alternative norm we use to study 
the growth of perturbations, given by eq. (\ref{loc_norm_2}), yields another expression for $g$,
\begin{equation}
\label{SFH_g_2}
g(t) = \frac{( \beta^2 + 1 )^2}
{\left [(\beta + q t )^2 + 1 \right ]^2}
\frac{1+\frac{2}{2-q}\, (\beta + q t )^2}
{1+\frac{2}{2-q}\, \beta^2},
\end{equation}
which is less tractable analytically, however,
along with eq. (\ref{SFH_g}) it recovers well the transient evolution of optimal perturbations
in case of $k_y, k_x\gg 1$.

\label{lastpage}

\end{document}